\documentclass[preprint,11pt]{elsarticle}

\usepackage{graphicx}
\usepackage{graphics}
\usepackage{epsfig}
\usepackage{pgfpages}
\usepackage{subfigure}
\usepackage{color}

\usepackage{listings}
\usepackage{psfrag}
\usepackage{booktabs}

\usepackage{amssymb}
\usepackage{mathrsfs}
\usepackage{mathrsfs}
\usepackage{mathrsfs}
\usepackage{amsfonts}
\usepackage{amsfonts}
\usepackage{amsfonts}
\usepackage{mathrsfs}
\usepackage{amsmath}
\usepackage{graphicx}
\usepackage{multirow}
\usepackage{caption2}
\usepackage{float}
\usepackage{booktabs}
\usepackage{array}

\newtheorem{Thm}{Theorem}
\newtheorem{Lemm}{Lemma}
\newtheorem{Prop}{Property}

\newtheorem{Assump}{Assumption}

\newcommand{\Min}{{\mathop{\mathrm{minimize}}}}
\newcommand{\diagnal}{{\mathbf{diag}}}
\newcommand{\nullspc}{{\mathbf{null}}}

\usepackage{subfigure}
\usepackage[colorlinks=trfue, linkcolor=black, citecolor=black, urlcolor=black, pdfborder={0 0 0}]{hyperref}
\usepackage[letterpaper,top=2in, bottom=1.5in, left=1in, right=1in]{geometry}
\numberwithin{equation}{section}

\journal{}

\begin{document}
\begin{frontmatter}



\title{A Fast Proximal Gradient Algorithm for Decentralized Composite Optimization over Directed Networks
 \tnoteref{t1}}
 \tnotetext[t1]{The work of J. Zeng is supported in part by the National Natural Science Foundation of China (Grants No. 61603162, 11401462).
 }

\author{Jinshan Zeng$^1$}

\author{Tao He$^1$}

\author{Mingwen Wang$^1$}

\address{1. College of Computer Information Engineering, Jiangxi Normal University, Nanchang, 330022, P R China.}

\begin{abstract}
This paper proposes a fast decentralized algorithm for solving a consensus optimization problem defined in a \emph{directed} networked multi-agent system, where the local objective functions have the smooth+nonsmooth composite form, and are possibly \emph{nonconvex}. Examples of such problems include decentralized compressed sensing and constrained quadratic programming problems, as well as many decentralized regularization problems.
We extend the existing algorithms PG-EXTRA and ExtraPush to a new algorithm \emph{PG-ExtraPush} for composite consensus optimization over a \emph{directed} network. This algorithm takes advantage of the proximity operator like in PG-EXTRA to deal with the nonsmooth term, and employs the push-sum protocol like in ExtraPush to tackle the bias introduced by the directed network. With a proper step size, we show that PG-ExtraPush converges to an optimal solution at a linear rate \footnote{In this paper, we use the notion of R-linear rate, i.e., a sequence $\{x^t\}$ converging to $x^*$ at an R-linear rate means that $\|x^t-x^*\|\leq C\rho^t$ for some constants $C>0$ and $\rho \in (0,1)$.} under some regular assumptions. We conduct a series of numerical experiments to show the effectiveness of the proposed algorithm. Specifically, with a proper step size, PG-ExtraPush performs linear rates in most of cases, even in some nonconvex cases, and is significantly faster than Subgradient-Push, even if the latter uses a hand-optimized step size. The established theoretical results are also verified by the numerical results.

\end{abstract}

\end{frontmatter}
\section{Introduction}
We consider the following consensus optimization problem defined on a \emph{directed}, strongly connected network of $n$ agents:
\begin{align}
\label{Eq:multi-agentOPT}
&\mathop{\Min}_{x\in \mathbb{R}^p} f(x) \triangleq \sum_{i=1}^n f_i(x),\nonumber\\
& \text{where} \quad f_i(x) = s_i(x)+ r_i(x),
\end{align}
and for every agent $i$, $f_i$ is a proper, coercive and possibly nonconvex function only known to the agent, $s_i$ is a smooth function, $r_i$ is generally nonsmooth and possibly nonconvex. We say that the objective has the smooth+nonsmooth composite structure.

The smooth+nonsmooth structure of the local objective arises in a large number of signal processing, statistical inference, and machine learning problems. Specific examples include: (i) the geometric median problem in which $s_i$ vanishes and $r_i$ is the $\ell_2$-norm \cite{Lopuhaa-GM1991}; (ii) the compressed sensing problem, where $s_i$ is the data-fidelity term, which is often differentiable, and $r_i$ is a sparsity-promoting regularizer such as the $\ell_q$ (quasi)-norm with $0\leq q \leq 1$ \cite{Ling-DCS2010}, \cite{Mateos-DistSparLinearReg2010}; (iii) optimization problems with per-agent constraints, where $s_i$ is a differentiable objective function of agent $i$ and $r_i$ is the indicator function of the constraint set of agent $i$, that is, $r_i(x)=0$ if $x$ satisfies the constraint and $\infty$ otherwise \cite{Chang-InexactADMM2015}, \cite{Lee-DistRandProj2013}.

For a stationary network with bi-directional communication, the existing algorithms include the primal-dual domain methods such as the decentralized alternating direction method of multipliers (DADMM) \cite{Schizas-DADMM2008,Shi-DADMM2014}, and the primal domain methods including the distributed subgradient method (DSM) \cite{Nedic-Subgradient2009}. Both algorithms do not take advantage of the smooth+nonsmooth structure. While the algorithms that consider smooth+nonsmooth objectives in the form of \eqref{Eq:multi-agentOPT} include the following primal-domain methods: the (fast) distributed proximal gradient method (DPGM) \cite{Chen-DPGM2012}, the proximal decentralized gradient descent method (Prox-DGD) \cite{Zeng-DGD2016}, the distributed iterative soft thresholding algorithm (DISTA) \cite{Ravazzi-DISTA2015}, proximal gradient exact first-order algorithm (PG-EXTRA) \cite{Shi-PGEXTRA2015}. All these primal-domain methods consist of a gradient step for the smooth part and a proximal step for the nonsmooth part. Different from DPGM, Prox-DGD and DISTA, PG-EXTRA as an extension of EXTRA \cite{Yin-EXTRA2015}  has two interlaced sequences of iterates, whereas the proximal-gradient method just inherits the sequence of iterates in the gradient method.

This paper focuses on a \emph{directed} network with \emph{directional} communication, which is pioneered by the works {\cite{Tsianos2012a-Alg,Tsianos2012b-Theory,Tsianos2013-PhdThsis}}. When communication is bi-directional, algorithms can use a symmetric and doubly-stochastic mixing matrix to obtain a consensual solution; however, once the communication is directional, the mixing matrix becomes generally asymmetric and only column-stochastic.
In the column-stochastic setting, the push-sum protocol \cite{Kempe-Push-sumw2003} can be used to obtain a stationary distribution for the mixing matrix. Some recent decentralized algorithms over a directed network include Subgradient-Push \cite{Nedic-SubgradientPush2015}, ExtraPush \cite{Zeng-ExtraPush2016} (also called DEXTRA in \cite{Xi-Khan2015}) and Push-DIGing \cite{Nedic-DIGing2016}.
The best rate of Subgradient-Push in the general convex case is $O(\ln t/\sqrt{t})$, where $t$ is the iteration number, and both ExtraPush and Push-DIGing perform linearly convergent in the strongly convex case. However, all of these algorithms do not consider the smooth+nonsmooth structure as well as the nonconvex case as defined in problem \eqref{Eq:multi-agentOPT}.

In this paper, we extend the algorithms PG-EXTRA and ExtraPush to the composite consensus optimization problem with the smooth+nonsmooth structure, and establish the convergence and linear convergence rate of the proposed PG-ExtraPush algorithm. At each iteration, each agent locally computes a gradient of the smooth part of its objective and a proximal map of the nonsmooth part, and exchanges information with its neighbors, then uses the push-sum protocol \cite{Kempe-Push-sumw2003} to achieve the consensus. When the network is undirected, the proposed PG-ExtraPush reduces to PG-EXTRA, and when $r_i\equiv 0$, PG-ExtraPush reduces to ExtraPush \cite{Zeng-ExtraPush2016}. If the smooth part of objective is Lipschitz differentiable and quasi-strongly convex and the nonsmooth part is convex with bounded subgradient (see Assumption \ref{Assump:ObjFun}), we prove that with a proper step size, the proposed algorithm converges to an optimal solution at a linear rate. We provide a series of numerical experiments including three convex cases and one nonconvex case, to show the effectiveness of the proposed algorithm.
Specifically, when applied to the convex cases, PG-ExtraPush performs the linear rates, and is significantly faster than Subgradient-Push, even if the latter uses a hand-optimized step size. While when applied to the nonconvex decentralized $\ell_q$ regularized least squares regression problems with $0\leq q<1$, it can be observed that the proposed algorithm performs an eventual linear convergence rate, that is, PG-ExtraPush performs a linear decay starting from a few iterations but not the initial iteration. This means that if we can fortunately get a good initial guess, the proposed algorithm PG-ExtraPush might decay linearly even in these nonconvex cases.

It should be pointed out that the extension from ExtraPush \cite{Zeng-ExtraPush2016} to PG-ExtraPush is non-trivial. The main differences between the proposed algorithm PG-ExtraPush and ExtraPush \cite{Zeng-ExtraPush2016} can be summarized as follows:
\begin{enumerate}
\item
{\bf On algorithm development.}
Clearly, PG-ExtraPush extends ExtraPush to handle nonsmooth objective terms. This extension is not the same as the extension from the gradient method to the proximal-gradient method, as well as the extension from EXTRA \cite{Yin-EXTRA2015} to PG-EXTRA \cite{Shi-PGEXTRA2015}. As the reader will see, PG-ExtraPush will have three interlaced sequences of iterates, whereas the proximal-gradient method just inherits of the sequence of iterates in the gradient method; and PG-ExtraPush uses the proximal maps of a sequence of transformed functions of $r_i$ associated with a positive weight sequence $\{{\bf w}^t\}$ essentially introduced by the directed graph,  while PG-EXTRA utilizes the proximity operator of $r_i$.

\item
{\bf On convergence analysis.} Although the convergence analysis of this paper is motivated by the existing analysis in \cite{Zeng-ExtraPush2016}, there are several new proof techniques.
The convergence of many existing algorithms like ExtraPush \cite{Zeng-ExtraPush2016} is established based on a similar inequality of \eqref{Eq:ConvThm} as presented in Theorem \ref{Thm:ConvThm}. However, we can not directly prove that such an inequality holds for all iterations of PG-ExtraPush. Instead, we can only establish the inequality \eqref{Eq:ConvThm} for a fixed iteration of PG-ExtraPush under the boundedness assumption of the previous two iterates. In order to establish the key inequality for all iterations, an induction technique is used as shown in the proof of Theorem \ref{Thm:ConvThm}. Moreover, the linear convergence rate of the proposed algorithm is established from the key inequality \eqref{Eq:ConvThm} via a recursive way. All of these are different from the convergence analysis in \cite{Zeng-ExtraPush2016}.
\end{enumerate}

The rest of paper is organized as follows. Section \ref{sc:problem} introduces the problem setup. Section \ref{sc:alg} develops the proposed algorithm. Section \ref{sc:converge} establishes its convergence and convergence rate. Section \ref{sc:experiment} presents our numerical results. We conclude this paper in Section \ref{sc:conclusion}.

\textbf{Notation:} Let ${\bf I}_n$ denote an identity matrix with the size $n\times n$. We use ${\bf 1}_{n} \in \mathbb{R}^{n}$ as a vector of all $1$'s. For any \emph{vector} $x$, we let $x_i$ denote its $i$th component and $\diagnal(x)$ denote the diagonal matrix generated by $x$. For any matrix $X$, $X^T$ denotes its transpose, $X_{ij}$ denotes its $(i,j)$th component, and $\|X\| \triangleq \sqrt{\langle X, X \rangle}=\sqrt{\sum_{i,j}X_{ij}^2}$ denotes its Frobenius norm. The largest and smallest eigenvalues of matrix $X$ are denoted as $\lambda_{\max}(X)$ and $\lambda_{\min}(X)$, respectively. For any matrix $B\in \mathbb{R}^{m\times n}$, $\nullspc(B) \triangleq \{x\in \mathbb{R}^n| Bx=0\}$ is the null space of $B$. \textbf{Given a matrix $B\in \mathbb{R}^{m\times n}$, by $Z\in \nullspc(B)$, we mean that each column of $Z$ lies in $\nullspc(B)$.}
The smallest \textit{nonzero} eigenvalue of a symmetric positive semidefinite matrix $X\neq {\bf 0}$ is denoted as $\tilde{\lambda}_{\min}(X)$, which is strictly positive. For any positive semidefinite matrix $G\in\mathbb{R}^{n\times n}$ (not necessarily symmetric in this paper), we use the notion $\|X\|_G^2 \triangleq \langle X, GX \rangle$ for a matrix $X \in \mathbb{R}^{n\times p}$.

\section{Problem reformulation}
\label{sc:problem}

\subsection{Network}

Consider a \emph{directed} network ${\cal G} = \{V,E\}$, where $V$ is the vertex set and $E$ is the edge set.
Any edge $(i,j)\in E$ represents a directed arc from node $i$ to node $j$.
The sets of in-neighbors and out-neighbors of  node $i$ are
\begin{align*}
{\cal{N}}_i^{\mathrm{in}} \triangleq \{j: (j,i) \in E\} \cup \{i\},\
 {\cal{N}}_i^{\mathrm{out}} \triangleq \{j: (i,j) \in E\} \cup \{i\},
\end{align*}
respectively.
Let $d_i \triangleq |{\cal{N}}_i^{\mathrm{out}}|$ be the out-degree of node $i$.
In ${\cal G}$, {each node $i$ can only send information to its out-neighbors, \emph{not} vice versa}.

\begin{figure}[!t]
\centering
\begin{minipage}[t][][c]{.48\textwidth}
\centering
\includegraphics[scale=0.8]{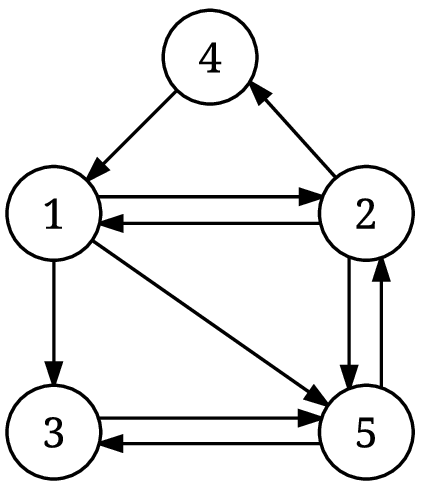}
\end{minipage}
\hspace{10pt}
\begin{minipage}[t][][b]{.48\textwidth}
$\displaystyle
A = \left(
\begin{array}{ccccc}
\frac{1}{4} &\frac{1}{4} & 0 & \frac{1}{2} & 0\\[3pt]
\frac{1}{4} &\frac{1}{4} & 0 & 0 & \frac{1}{3}\\[3pt]
\frac{1}{4}  &0 & \frac{1}{2} & 0 & \frac{1}{3} \\[3pt]
0 &\frac{1}{4} & 0 & \frac{1}{2} & 0\\[3pt]
\frac{1}{4} & \frac{1}{4} & \frac{1}{2} & 0 & \frac{1}{3}\\
\end{array}
\right)$
\end{minipage}
\caption{ A directed graph ${\cal G}$ (left) and its mixing matrix $A$ (right) \cite{Zeng-ExtraPush2016}.}
\label{Fig:Network}
\end{figure}

To illustrate a mixing matrix for a directed network, consider $A\in \mathbb{R}^{n\times n}$ where 
\begin{equation}
\label{Eq:MixingMatrixA}
\left\{
\begin{array}{ll}
A_{ij}>0, & \text{if}\ j\in {\cal N}_i^{\mathrm{in}}\\
A_{ij}=0, & \text{otherwise.}
\end{array}%
\right.
\end{equation}%
The entries $A_{ij}$ satisfy that, for each node $j$,  $\sum_{i\in V}A_{ij}=1$. An example is the following mixing matrix
\begin{equation}
\label{Eq:MixingMatrixA*}
A_{ij} =
\left\{
\begin{array}{ll}
1/d_j, & \text{if}\ j\in {\cal N}_i^{in}\\
0, & \text{otherwise}
\end{array}%
\right.,
\end{equation}%
$i,j=1,\ldots, n$, which is used in the Subgradient-Push method {\cite{Nedic-SubgradientPush2015}}.
See Fig. {\ref{Fig:Network}} for a directed graph ${\cal G}$  and an example of its mixing matrix $A$.
The matrix $A$ is column stochastic and  asymmetric in general.

\begin{Assump}
\label{Assump:conectivityofgraph}
The graph ${\cal G}$ is strongly connected.
\end{Assump}

\begin{Prop}
\label{Prop:A}
Under Assumption \ref{Assump:conectivityofgraph}, the following hold (parts (i) and (iv) are results in \cite[Corollary 2]{Nedic-SubgradientPush2015}, while parts (ii) and (iii) are results in \cite[Property 1]{Zeng-ExtraPush2016})
\begin{enumerate}
\item[(i)]
Let $A^t = \overbrace{A\times A \cdots A}^t$ for any $t\in \mathbb{N}$. Then
\begin{align}
\label{Eq:LimofAt}
A^t \rightarrow \phi {\bf 1}_n^T \ \ \mathrm{geometrically\  fast\  as }\ \ t \rightarrow \infty,
\end{align}
for some \emph{stationary distribution} vector $\phi$, i.e., $\phi_i\ge 0$ and $\sum_i^n \phi_i =1$.
More specifically, for all $i\in \{1,\ldots,n\}$, the entries $(A^t)_{ij}$ and $\phi_i$, there holds
\[
|(A^t)_{ij} -\phi_i| < C\gamma^t, \quad \forall j\in \{1,\ldots,n\},
\]
where $C=4$ and $\gamma = 1-\frac{1}{n^n}$.

\item[(ii)]
$\nullspc({\bf I}_n - \phi {\bf 1}_n^T) = \nullspc({\bf I}_n - A).$

\item[(iii)]
$A\phi = \phi.$

\item[(iv)]
The quantity $\xi \triangleq \inf_t \min_{1\leq i \leq n} (A^t{\bf 1}_n)_i \geq \frac{1}{n^n}>0.$
\end{enumerate}
\end{Prop}
Letting,
\begin{align}
\label{Eq:def-D}
D_{\infty} \triangleq n\diagnal(\phi), \quad D^t \triangleq \diagnal(A^t {\bf 1}_n),
\end{align}
and
\begin{align}
\label{Eq:def-d}
d^+ \triangleq \max_t \{\|D^t\|\},\  d^- \triangleq \max_t \{\|(D^t)^{-1}\|\},\  d_{\infty}^+ \triangleq \|D_{\infty}\|,\  d_{\infty}^- \triangleq \|D_{\infty}^{-1}\|.
\end{align}
Thus, by Property \ref{Prop:A}, there hold
\begin{align}
\label{Eq:rate-D-Dt}
\|D^t - D_{\infty}\| \leq nC\gamma^t,
\end{align}
\begin{align}
\label{Eq:rate-invD-invDt}
\|(D^t)^{-1} - (D_{\infty})^{-1}\| = \|(D^t)^{-1}(D_{\infty}-D^t)(D_{\infty})^{-1}\| \leq {d^-}{d_{\infty}^-}nC\gamma^t,
\end{align}
for any $t\in \mathbb{N}$.

\subsection{Problem with matrix notation}

Let $x_{(i)} \in \mathbb{R}^p$ denote the \emph{local copy} of $x$ at node $i$, and
$x_{(i)}^t$ denote its value at the $t$-th iteration.
Throughout the note, we  use the following equivalent form of the problem (\ref{Eq:multi-agentOPT}) using {local copies} of the  variable $x$:
\begin{align}\label{Eq:consensusProblem}
\Min_{\mathbf{x}}\  {\bf 1}_n^T {\bf f(x)} \triangleq \sum_{i=1}^n f_i(x_{(i)}),\nonumber\\
\mathrm{subject\  to}\ x_{(i)} = x_{(j)}, \ \forall (i,j)\in E,
\end{align}
where ${\bf 1}_n \in \mathbb{R}^n$ denotes the vector with all its entries equal to 1, ${\bf x}\in \mathbb{R}^{n\times p}$, ${\bf f(x)}\in \mathbb{R}^{n}$, ${\bf s(x)}\in \mathbb{R}^{n}$ and ${\bf r(x)}\in \mathbb{R}^{n}$ with
$$
{\bf x} \triangleq \left(
\begin{array}{ccc}
\mbox{---} &x^T_{(1)} & \mbox{---}\\
\mbox{---} &x^T_{(2)} & \mbox{---}\\
  &\vdots &  \\
\mbox{---} &x^T_{(n)} & \mbox{---}\\
\end{array}
\right)
,\quad
{\bf f(x)} \triangleq \left(
\begin{array}{c}
f_1(x_{(1)})\\
f_2(x_{(2)})\\
\vdots   \\
f_n(x_{(n)})\\
\end{array}
\right)
,
{\bf s(x)} \triangleq \left(
\begin{array}{c}
s_1(x_{(1)})\\
s_2(x_{(2)})\\
\vdots   \\
s_n(x_{(n)})\\
\end{array}
\right)
,\quad
{\bf r(x)} \triangleq \left(
\begin{array}{c}
r_1(x_{(1)})\\
r_2(x_{(2)})\\
\vdots   \\
r_n(x_{(n)})\\
\end{array}
\right)
.
$$
In addition, the gradient of ${\bf s}(\bf x)$ is
$$
{\bf \nabla s(x)} \triangleq \left(
\begin{array}{ccc}
\mbox{---} &\nabla s_1(x_{(1)})^T & \mbox{---}\\
\mbox{---} &\nabla s_2(x_{(2)})^T & \mbox{---}\\
  &\vdots &  \\
\mbox{---} &\nabla s_n(x_{(n)})^T & \mbox{---}\\
\end{array}
\right)
\in \mathbb{R}^{n\times p},
$$
and a subgradient of ${\bf r}(\bf x)$ is
$$
{\bf \tilde{\nabla} r(x)} \triangleq \left(
\begin{array}{ccc}
\mbox{---} &\tilde{\nabla} r_1(x_{(1)})^T & \mbox{---}\\
\mbox{---} &\tilde{\nabla} r_2(x_{(2)})^T & \mbox{---}\\
  &\vdots &  \\
\mbox{---} &\tilde{\nabla} r_n(x_{(n)})^T & \mbox{---}\\
\end{array}
\right)
\in \mathbb{R}^{n\times p}.
$$
The $i$th rows of the above matrices $\mathbf{x}$, $\nabla \mathbf{s}(\mathbf{x})$ and ${\bf \tilde{\nabla} r(x)}$, and vector
${\bf s(x)}$, correspond to agent $i$. For simplicity, one can treat $p=1$ throughout this paper.
To deal with the nonsmooth part, given a parameter $\alpha>0$, we introduce the proximity operator $\mathrm{prox}_{\alpha  r_i}$ associated with $r_i$ as follows
\begin{align}
\label{Eq:prox}
\mathrm{prox}_{\alpha r_i}(z) = \mathop{\mathrm{argmin}}_{u\in \mathbb{R}^p} \left\{r_i(u) + \frac{\|u-z\|_2^2}{2\alpha}\right\}.
\end{align}
For any ${\bf z}\in \mathbb{R}^{n\times p}$, define
$$\mathrm{Prox}_{\alpha {\bf r}} ({\bf z}) =
\left(
\begin{array}{c}
\mathrm{prox}_{\alpha r_1}(z_{(1)})\\
\mathrm{prox}_{\alpha r_2}(z_{(2)})\\
\vdots\\
\mathrm{prox}_{\alpha r_n}(z_{(n)})
\end{array}
\right).$$

\section{Development of Algorithm}
\label{sc:alg}

\subsection{Proposed Algorithm: PG-ExtraPush}

The proposed algorithm PG-ExtraPush extends PG-EXTRA and ExtraPush to composite (smooth+nonsmooth) consensus optimization problem. Given a sequence of $n$-dimensional positive vectors $\{{\bf w}^t\}_{t\in \mathbb{N}}$, we define a sequence of functions
\[
{\bf r}^t({\bf x}) \triangleq \diagnal({\bf w}^t){\bf r}(\diagnal({\bf w}^t)^{-1}{\bf x}), \ \forall \ {\bf x}\in \mathbb{R}^{n\times p}, \ t \in \mathbb{N}.
\]

Let $\bar{A} \triangleq \frac{A+{\bf I}_n}{2}$.
Specifically, the proposed algorithm can be described as follows: for all agents $i=1,\ldots,n$, set arbitrary $z_{(i)}^0 \in \mathbb{R}^p$, $w_i^0 = 1$, $x_{(i)}^0 = z_{(i)}^0$; $z_{(i)}^{1/2} = \sum_{j=1}^n A_{ij} z_{(j)}^0 - \alpha \nabla s_i(z_{(i)}^0)$, $w_i^1 = \sum_{j=1}^n A_{ij}w_j^0$, $z_{(i)}^1 = \mathrm{prox}_{\alpha r_i^1}(z_{(i)}^{1/2})$, $x_{(i)}^1 = \frac{z_{(i)}^1}{w_i^1}$. For $t=1,2\ldots,$ perform
\begin{equation}
\label{Eq:PG-ExtraPush_vectorform}
\left\{
\begin{array}{l}
z_{(i)}^{t+1/2} = \sum_{j=1}^n A_{ij}z_{(j)}^{t} + z_{(i)}^{t-1/2} - \sum_{j=1}^n \bar{A}_{ij}z_{(j)}^{t-1} - \alpha (\nabla s_i(x_{(i)}^{t}) - \nabla s_i(x_{(i)}^{t-1})),\\
w_i^{t+1} = \sum_{j=1}^n A_{ij}w_j^{t},\\
z_{(i)}^{t+1} = \mathrm{prox}_{\alpha r_i^{t+1}}(z_{(i)}^{t+1/2}),\\
x_{(i)}^{t+1} = \frac{z_{(i)}^{t+1}}{w_i^{t+1}}.
\end{array}%
\right.
\end{equation}%

The matrix form of the algorithm can be described as follows: set arbitrary ${\bf z}^0 \in \mathbb{R}^{n\times p}$, ${\bf w}^0 = {\bf 1}_n$, ${\bf x}^0 = {\bf z}^0$; ${\bf z}^{1/2} = A {\bf z}^0 - \alpha \nabla {\bf s}({\bf z}^0)$, ${\bf w}^1 = A{\bf w}^0$, ${\bf z}^1 = \mathrm{Prox}_{\alpha {\bf r}^1}({\bf z}^{1/2})$, ${\bf x}^1 = \diagnal{({\bf w}^1)}^{-1}{\bf z}^1$. For $t=1,2,\ldots,$ perform
\begin{equation}
\label{Eq:PG-ExtraPush_matrixform}
\left\{
\begin{array}{l}
{\bf z}^{t+1/2} = A{\bf z}^{t} + {\bf z}^{t-1/2} - \bar{A}{\bf z}^{t-1} - \alpha (\nabla {\bf s}({\bf x}^{t}) - \nabla {\bf s}({\bf x}^{t-1})),\\
{\bf w}^{t+1} = A {\bf w}^{t},\\
{\bf z}^{t+1} =  \mathrm{Prox}_{\alpha {\bf r}^{t+1}}({\bf z}^{t+1/2}),\\
{\bf x}^{t+1} = \diagnal{({\bf w}^{t+1})}^{-1}{\bf z}^{t+1}.
\end{array}%
\right.
\end{equation}%

By the definition of the proximal operator and the definition of function ${\bf r}^{t}$, the PG-ExtraPush iteration \eqref{Eq:PG-ExtraPush_matrixform} implies
\begin{align}
\label{Eq:z-update0}
{\bf z}^{t+1}
= {\bar A}{\bf z}^t + {\bar A}({\bf z}^t - {\bf z}^{t-1}) - \alpha (\nabla {\bf s}({\bf x}^t)-\nabla {\bf s}({\bf x}^{t-1}))
- \alpha (\tilde{\nabla} {\bf r}({\bf x}^{t+1})-\tilde{\nabla} {\bf r}({\bf x}^{t})),
\end{align}
for $t=1,2,\ldots.$

\subsection{Special Cases: PG-EXTRA, ExtraPush and P-ExtraPush}

When the network is undirected, then the weight sequence ${\bf w}^t \equiv {\bf 1}_n$, thus, the function ${\bf r}^t \equiv {\bf r}$ and the sequence ${\bf x}^t = {\bf z}^t$. Therefore, PG-ExtraPush reduces to PG-EXTRA \cite{Shi-PGEXTRA2015}, a recent algorithm for composite consensus optimization over undirected networks.

When the possibly-nondifferentiable term ${\bf r} \equiv 0,$ we have ${\bf z}^1 = {\bf z}^{1/2}$, and thus, ${\bf z}^1 = A {\bf z}^0 - \alpha \nabla {\bf s}({\bf z}^0).$
In the third update of \eqref{Eq:PG-ExtraPush_matrixform}, ${\bf z}^{t+1} = {\bf z}^{t+1/2}$, and thus
\begin{equation}
\label{Eq:PG-ExtraPushzt}
{\bf z}^{t+1} = A{\bf z}^{t} + {\bf z}^{t} - \bar{A}{\bf z}^{t-1} - \alpha (\nabla {\bf s}({\bf x}^{t}) - \nabla {\bf s}({\bf x}^{t-1})).
\end{equation}
With these, in this case, PG-ExtraPush reduces to ExtraPush \cite{Zeng-ExtraPush2016}, a recent algorithm for decentralized differentiable optimization over directed networks.

When the differentiable term ${\bf s}\equiv 0$, PG-ExtraPush reduces to P-ExtraPush by removing all gradient computation, which is given as follows:
set arbitrary ${\bf z}^0 \in \mathbb{R}^{n\times p}$, ${\bf w}^0 = {\bf 1}_n$, ${\bf x}^0 = {\bf z}^0$; ${\bf z}^{1/2} = A {\bf z}^0 $, ${\bf w}^1 = A{\bf w}^0$, ${\bf z}^1 = \mathrm{Prox}_{\alpha {\bf r}^1}({\bf z}^{1/2})$, ${\bf x}^1 = \diagnal{({\bf w}^1)}^{-1}{\bf z}^1$. For $t=1,2,\ldots,$ perform
\begin{equation}
\label{Eq:P-ExtraPush_matrixform}
\left\{
\begin{array}{l}
{\bf z}^{t+1/2} = A{\bf z}^{t} + {\bf z}^{t-1/2} - \bar{A}{\bf z}^{t-1},\\
{\bf w}^{t+1} = A {\bf w}^{t},\\
{\bf z}^{t+1} =  \mathrm{Prox}_{\alpha {\bf r}^{t+1}}({\bf z}^{t+1/2}),\\
{\bf x}^{t+1} = \diagnal{({\bf w}^{t+1})}^{-1}{\bf z}^{t+1}.
\end{array}%
\right.
\end{equation}%

\section{Convergence Analysis}
\label{sc:converge}

In this section, we analyze the convergence of the proposed algorithm.

\subsection{Assumptions}
In this subsection, we presents the main assumptions. Besides the strongly connected assumption on the directed graph, we still need the following assumptions.

\begin{Assump}
\label{Assump:ExisOptSolution}
{\bf (existence of solution)}
Let ${\cal X}^*$ be the optimal solution set of problem \eqref{Eq:multi-agentOPT}, and   assume that  ${\cal X}^*$ is nonempty.
\end{Assump}

\begin{Assump}
\label{Assump:ObjFun}
For each agent $i$, its objective functions $s_i$ and $r_i$ satisfy the following:
\begin{enumerate}
\item[(i)] {\bf (Lipschitz differentiability)}
$s_i$ is  differentiable, and its gradient $\nabla s_i$ is $L_i$-Lipschitz continuous, i.e.,
$\|\nabla s_i(x)-\nabla s_i(y)\|\leq L_i \|x-y\|, \forall x,y\in \mathbb{R}^p$;

\item[(ii)] {\bf (quasi-strong convexity)}
$s_i$ is quasi-strongly convex, and there exists a positive constant $\mu_i$
such that $\mu_i \|x^*-y\|^2 \leq \langle \nabla s_i(x^*)-\nabla s_i(y), x^*-y \rangle$ for any $y\in \mathbb{R}^p$
and some optimal value $x^* \in {\cal X}^*$.

\item[(iii)]{\bf (bounded subgradient)}
$r_i$ is convex and $\tilde{\nabla}r_i(x)$ is uniformly bounded by some constant $B_{r_i}$, i.e., $\|\tilde{\nabla}r_i(x)\| \leq B_{r_i}$ for any $x\in \mathbb{R}^p$.
\end{enumerate}
\end{Assump}

Following Assumption {\ref{Assump:ObjFun}}, there hold for any ${\bf x}, {\bf y} \in \mathbb{R}^{n\times p}$ and some ${\bf x}^* \equiv {\bf 1}_n {(x^*)}^T$
\begin{align}
&\|\nabla {\bf s}({\bf x}) - \nabla {\bf s}({\bf y})\| \leq L_s \|{\bf x}-{\bf y}\|,\label{Eq:LipC-Grad}\\
&\mu_s \|{\bf x}^*-{\bf y}\|^2 \leq \langle \nabla {\bf s}({\bf x}^*)-\nabla {\bf s}({\bf y}), {\bf x}^*-{\bf y}\rangle,  \label{Eq:StrongCVX}\\
& \|\tilde{\nabla}{\bf r}({\bf x})\| \leq B_r, \label{Eq:boundedsubgrad}
\end{align}
where the constants $L_s \triangleq \max_i L_i$, $\mu_s \triangleq \min_i \mu_i$, and $B_r \triangleq \sum_{i=1}^n B_{r_i}$. The Lipschitz differentiable and strongly convex assumptions (Assumption \ref{Assump:ObjFun}(i), (ii)) are generally necessary to derive the linear convergence of decentralized algorithms such as in \cite{Yin-EXTRA2015,Zeng-ExtraPush2016}. While the bounded subgradient assumption (Assumption \ref{Assump:ObjFun}(iii)) is a regular assumption in the convergence analysis of decentralized algorithms like in \cite{Chen-DPGM2012,Nedic-Subgradient2009}. There are many functions satisfy Assumption \ref{Assump:ObjFun}(iii) such as the $\ell_1$ norm and Huber function, which are widely used in machine learning and compressed sensing. Actually, according to the latter proof of Theorem \ref{Thm:ConvThm}, the requirement of the uniformly bounded subgradient of $r_i$ can be relaxed to the boundedness of $\{\tilde{\nabla} r_i({\bf x}_{i}^t)\}_{t\in \mathbb{N}}$.

\begin{Assump}
\label{Assump:WeightingMatrices}
{\bf (positive definiteness)}
$D_{\infty}^{-1}\bar{A}+\bar{A}^TD_{\infty}^{-1} \succ 0$.
\end{Assump}

By noticing $D_{\infty}^{-1}\bar{A}+\bar{A}^TD_{\infty}^{-1} = D_{\infty}^{-1/2}(D_{\infty}^{-1/2}\bar{A}D_{\infty}^{1/2} + D_{\infty}^{1/2}\bar{A}^T D_{\infty}^{-1/2})D_{\infty}^{-1/2}$, we can guarantee the positive definiteness of $D_{\infty}^{-1}\bar{A}+\bar{A}^TD_{\infty}^{-1}$ by ensuring the matrix $\bar{A}+\bar{A}^T$ to be positive definite.
Note that $\bar{A}_{ii} > \sum_{j\neq i} \bar{A}_{ij}$ for each $i$, which means that $\bar{A}$ is strictly column-diagonal dominant.
To ensure the positive definiteness of $\bar{A}+\bar{A}^T$, each node $j$ can be ``selfish" and take a sufficiently large $A_{jj}$.

\subsection{Main Results}

In this subsection, we first develop the first-order optimality conditions for the problem \eqref{Eq:consensusProblem}
and then establish the convergence and convergence rate of PG-ExtraPush under the above assumptions.

\begin{Thm}[first-order optimality conditions]
\label{Thm:1stOrderOptCond}
Let Assumption \ref{Assump:conectivityofgraph} hold.
Then ${\bf x}^*$ is consensual and $x^*_{(1)} \equiv x^*_{(2)} \equiv \cdots \equiv x^*_{(n)}$ is an optimal solution of (\ref{Eq:multi-agentOPT})
if and only if, for some $\alpha>0$, there exist ${\bf z}^* \in \nullspc({\bf I}_n - A)$ and ${\bf y}^* \in \nullspc({\bf 1}_n^T)$
such that the following conditions hold
\begin{equation}
\label{Eq:1stOrderOpt}
\left\{
\begin{array}{l}
{\bf y}^* + \alpha (\nabla {\bf s}({\bf x}^*)+\tilde{\nabla}{\bf r}({\bf x}^*)) = 0,\\
{\bf x}^* = D_{\infty}^{-1} {\bf z}^*.
\end{array}%
\right.
\end{equation}%
(We let ${\cal L}^*$ denote the set of triples $({\bf z}^*, {\bf y}^*, {\bf x}^*)$ satisfying the above conditions.)
\end{Thm}

Theorem \ref{Thm:1stOrderOptCond} gives some equivalent conditions to characterize the optimal solution of the original optimization problem \eqref{Eq:multi-agentOPT}. Based on Theorem \ref{Thm:1stOrderOptCond}, we give the following subsequence convergence of PG-ExtraPush under the boundedness assumption of sequence $\{{\bf z}^t\}$.
\begin{Thm}[subsequence convergence under boundedness assumption]
\label{Thm:LimitPoint_Optimality}
Let Assumption \ref{Assump:conectivityofgraph}, Assumption \ref{Assump:ExisOptSolution}, and Assumption \ref{Assump:ObjFun}(i), (iii) hold. Let $\{({\bf z}^t, {\bf x}^t, {\bf w}^t)\}$ be any sequence generated by PG-ExtraPush {\eqref{Eq:PG-ExtraPush_matrixform}}.
Define ${\bf y}^t \triangleq \sum_{k=0}^t (\bar{A}-A){\bf z}^k$.
Suppose that $\{{\bf z}^t\}$ is bounded. Then, there exists a convergent subsequence of $\{({\bf z}^t, {\bf y}^t, {\bf x}^t)\}$, and any limit point of $\{({\bf z}^t, {\bf y}^t, {\bf x}^t)\}$, denoted by $({\bf z}^*, {\bf y}^*, {\bf x}^*)$, satisfies the optimality conditions (\ref{Eq:1stOrderOpt}).
\end{Thm}



From Theorem \ref{Thm:LimitPoint_Optimality}, if $\{{\bf z}^t\}$ is bounded, then both $\{{\bf x}^t\}$ and $\{{\bf y}^t\}$ are also bounded, and thus, there exists a convergent subsequence, and any limit point is an optimal solution of the original consensus optimization problem. However, it is generally difficult to verify the boundedness of $\{{\bf z}^t\}$. To guarantee this, we may need more assumptions on the objective functions such as the strong convexity of the smooth term.
In the following, we present the convergence and linear convergence rate of PG-ExtraPush under these additional assumptions. Before presenting the main result, we introduce the following notations. For each $t$, introducing ${\bf u}^t = \sum_{k=0}^t {\bf z}^k$,
then similar to \eqref{Eq:Var1ofPG-EXTRAPush}, the PG-ExtraPush iteration {\eqref{Eq:PG-ExtraPush_matrixform}} reduces to
\begin{equation}
\left\{
\begin{array}{l}
\bar{A}{\bf z}^{t+1} = \bar{A}{\bf z}^t - \alpha \tilde{\nabla} {\bf r}({\bf x}^{t+1}) -\alpha \nabla {\bf f}({\bf x}^t) - (\bar{A}-A){\bf u}^{t+1}\\
{\bf u}^{t+1} = {\bf u}^t + {\bf z}^{t+1}\\
{\bf x}^{t+1} = (D^{t+1})^{-1} {\bf z}^{t+1}.
\end{array}%
\right.
\label{Eq:PG-ExtraPush_var2}
\end{equation}
Let $({\bf z}^*, {\bf y}^*, {\bf x}^*) \in {\cal L}^*$, where ${\bf x}^*$ has been specified in {\eqref{Eq:StrongCVX}}.
Let ${\bf u}^*$ be any matrix that satisfies $(\bar{A}-A){\bf u}^* = {\bf y}^*$.
For simplicity, we introduce
\begin{equation}
\label{Eq:MetricForm}
{\bf v}^t = \left(
\begin{array}{c}
{\bf z}^t\\
{\bf u}^t
\end{array}
\right),~
{\bf v}^* = \left(
\begin{array}{c}
{\bf z}^*\\
{\bf u}^*
\end{array}
\right),~
G = \left(
\begin{array}{cc}
N^T & {\bf 0}\\
{\bf 0} & M
\end{array}
\right),
~S = \left(
\begin{array}{cc}
{\bf 0} & M\\
 -M^T & {\bf 0}
\end{array}
\right),
\end{equation}
where $N = D_{\infty}^{-1}\bar{A}$, $M =  D_{\infty}^{-1}(\bar{A}-A)$.
Let ${\bf e}^t \triangleq
\left(\begin{array}{c}
D_{\infty}^{-1}(\tilde{\nabla} {\bf r}({\bf x}^{t+1}) + \nabla {\bf f}({\bf x}^t))\\
0
\end{array}
\right).
$
By {\eqref{Eq:PG-ExtraPush_var2}} and {\eqref{Eq:MetricForm}}, the PG-ExtraPush iteration {\eqref{Eq:PG-ExtraPush_matrixform}} implies
\begin{align}
\label{Eq:PG-ExtraPush_var3}
G^T({\bf v}^{t+1} - {\bf v}^t) = -S{\bf v}^{t+1} -\alpha {\bf e}^t.
\end{align}
According to \cite{Zeng-ExtraPush2016}, both $M+M^T$ and $G+G^T$ are positive semidefinite,
and the following property holds
\[
\|x\|_{G}^2 = \frac{1}{2}\|x\|_{G+G^T}^2\geq 0, \quad\forall x\in \mathbb{R}^{2n}.
\]

Let $c_1 = \frac{\lambda_{\max}(MM^T)}{\tilde{\lambda}_{\min}(M^TM)}$,
$c_2 = \frac{\lambda_{\max}(\frac{M+M^T}{2})}{\tilde{\lambda}_{\min}(M^TM)}$, and
$c_3 = \lambda_{\max}(NN^T) + 3c_1\lambda_{\max}(N^TN)$.
Let $\bar{L} = {d_{\infty}^-}{d^-}L_s$.
Let $\Delta_1 = (\bar{\mu} -\frac{\bar{\eta}}{2})^2 - 6c_1{\bar L}^2$,
and $\Delta_2 = \frac{{\bar L}^4}{4\bar{\eta}^2} -3c_1{\bar L}^2\sigma (c_3\sigma - \lambda_{\min}(N^T+N))$ for some appropriate tunable parameters $\bar{\eta}$ and $\sigma$.
Then we describe our main result as follows.

\begin{Thm}[linear convergence rate]
\label{Thm:ConvThm}
Let Assumptions {\ref{Assump:conectivityofgraph}}-{\ref{Assump:WeightingMatrices}} hold.
If the step size parameter $\alpha$ satisfies
\begin{align}
\label{Eq:Condonalpha*}
\frac{{\bar \mu} - \frac{\bar{\eta}}{2} - \sqrt{\Delta_1}}{3c_1{\bar L}^2\sigma} < \alpha < \min\left\{\frac{{\bar \mu} - \frac{\bar{\eta}}{2} + \sqrt{\Delta_1}}{3c_1{\bar L}^2\sigma}, \frac{-\frac{{\bar L}^2}{2\bar{\eta}}+ \sqrt{\Delta_2}}{3c_1{\bar L}^2 \sigma}\right\}
\end{align}
for some appropriate $\bar{\eta}$ and $\sigma$ as specified in {\eqref{Eq:Condoneta}} and {\eqref{Eq:Condonsigma}}, respectively, then the sequence $\{{\bf v}^t\}$ defined in {\eqref{Eq:MetricForm}} satisfies
\begin{equation}
\label{Eq:ConvThm}
\|{\bf v}^{t}-{\bf v}^*\|_G^2 \geq (1+\delta)\|{\bf v}^{t+1}-{\bf v}^*\|_G^2 - \Gamma_0 \gamma^t,
\end{equation}
for  $\delta>0$ obeying
\begin{align}
\label{Eq:Condondelta}
0<\delta \leq
\min\bigg\{ \frac{-\frac{1}{\sigma} +(\bar{\mu} - \frac{\bar{\eta}}{2})\alpha - \frac{3}{2}c_1{\bar L}^2\sigma \alpha^2}{\lambda_{\max}(\frac{N+N^T}{2})+ 3c_2\alpha^2{\bar L}^2},
\frac{\lambda_{\min}(\frac{N^T+N}{2}) - \frac{c_3\sigma}{2} - \frac{{\bar L}^2\alpha}{2\bar{\eta}} - \frac{3}{2}c_1{\bar L}^2\sigma \alpha^2}{3c_2(\lambda_{\max}(N^TN)+\alpha^2{\bar L}^2)}\bigg\},
\end{align}
and a constant $\Gamma_0$ as specified in \eqref{Eq:def-gamma0}. Furthermore, \eqref{Eq:ConvThm} implies that the sequence $\{{\bf x}^t\}$ converges to an optimal solution ${\bf x}^*$ at a linear rate, i.e.,
\begin{align}
\label{Eq:LinConvThm}
\|{\bf x}^t - {\bf x}^*\| \leq \Gamma (\sqrt{\rho})^t,
\end{align}
for some $\max\{\frac{1}{1+\delta},\gamma\}<\rho <1$,
where $\Gamma$ is specified in \eqref{Eq:def-gamma}.
\end{Thm}

From this theorem, the sequence $\{{\bf x}^t\}$ converges to ${\bf x}^*$ at a linear rate.
By the definition of ${\bf v}^*$ in {\eqref{Eq:MetricForm}}, ${\bf v}^*$ is indeed defined by some optimal value $({\bf z}^*, {\bf y}^*, {\bf x}^*)$.
Roughly speaking, bigger $\delta$ means faster convergence rate. As specified in Theorem {\ref{Thm:ConvThm}}, $\delta$ is affected by many factors.
Generally, $\delta$ decreases with respect to both $\lambda_{\max}(\frac{N+N^T}{2})$ and $\lambda_{\max}(N^TN)$,
which potentially implies that if all nodes are more ``selfish", that is, they hold more information for themselves than sending  to their out-neighbors.
Consequently, the information mixing speed of the network will get smaller, and thus the convergence of  PG-ExtraPush becomes slower.
Therefore, we suggest a more democratic rule (such as the matrix $A$ specified in {\eqref{Eq:MixingMatrixA*}})  for faster convergence in practice.
To ensure $\delta>0$, it requires that the step size $\alpha$  lie in an appropriate interval.
It should be pointed out that the condition \eqref{Eq:Condonalpha*} on $\alpha$ is sufficiently, not necessary, for the linear convergence of PG-ExtraPush. In fact, in the next section, it can be observed that PG-ExtraPush algorithm  converges under small values of $\alpha$. In general, a smaller $\alpha$ implies a slower rate of convergence. According to the definition of $\Delta_2$ and the condition \eqref{Eq:Condonsigma} on $\sigma$, the upper bound of step size $\alpha$ in \eqref{Eq:Condonalpha*} implies that
\[
\alpha < \frac{-\frac{{\bar L}^2}{2\bar{\eta}}+ \sqrt{\Delta_2}}{3c_1{\bar L}^2 \sigma} \leq \frac{\sqrt{3c_1\sigma(\lambda_{\min}(N^T+N)-c_3\sigma)}\bar{L}}{3c_1\sigma \bar{L}^2}
= \sqrt{\frac{\lambda_{\min}(N^T+N)-c_3\sigma}{3c_1\sigma}} \times \frac{1}{\bar{L}}.
\]
It can be observed from the above relation that the upper bound of step size is inversely proportional to the Lipschitz constant of $\nabla {\bf s}$, which is a regular condition for the convergence of a proximal-type algorithm.

\subsection{Proofs}
In this subsection, we provide the proofs of Theorems \ref{Thm:1stOrderOptCond}, \ref{Thm:LimitPoint_Optimality} and \ref{Thm:ConvThm}.

\subsubsection{Proof of Theorem \ref{Thm:1stOrderOptCond}}
{\bf Proof.}
Assume that ${\bf x}^*$ is consensual and $x^*_{(1)} \equiv x^*_{(2)} \equiv \cdots \equiv x^*_{(n)}$ is optimal.
Let ${\bf z}^* = n \diagnal(\phi) {\bf x}^* = n( \phi x^{*T}_{(1)}).$
Then $\phi {\bf 1}_n^T {\bf z}^* = \phi {\bf 1}_n^T n \phi x^{*T}_{(1)} = n \phi x^{*T}_{(1)}= {\bf z}^*.$
It implies that ${\bf z}^* \in \nullspc({\bf I} - \phi {\bf 1}_n^T)$.
By Property {\ref{Prop:A}}(ii), it follows that ${\bf z}^* \in \nullspc({\bf I}_n - A)$.
Moreover, letting ${\bf y}^* = -\alpha (\nabla {\bf s}({\bf x}^*)+\tilde{\nabla} {\bf r}({\bf x}^*)),$ it holds that ${\bf 1}_n^T {\bf y}^* = -\alpha {\bf 1}_n^T (\nabla {\bf s}({\bf x}^*)+\tilde{\nabla} {\bf r}({\bf x}^*)) = 0$, that is, ${\bf y}^* \in \nullspc({\bf 1}_n^T)$.

On the other hand, assume (\ref{Eq:1stOrderOpt}) holds. By Property {\ref{Prop:A}}(ii), it follows that ${\bf z}^* = \phi {\bf 1}_n^T {\bf z}^*.$
Plugging ${\bf x}^* = D_{\infty}^{-1} {\bf z}^*$ gives ${\bf x}^* = \frac{1}{n}{\bf 1}_n{\bf 1}_n^T {\bf z}^*,$
which implies that ${\bf x}^*$ is consensual. Moreover, by ${\bf y}^* + \alpha (\nabla {\bf s}({\bf x}^*)+\tilde{\nabla} {\bf r}({\bf x}^*)) = 0$
and ${\bf y}^* \in \nullspc({\bf 1}_n^T)$, it holds ${\bf 1}_n^T (\nabla {\bf s}({\bf x}^*)+\tilde{\nabla} {\bf r}({\bf x}^*)) = -\frac{1}{\alpha}{\bf 1}_n^T{\bf y}^* =0$,
which implies that ${\bf x}^*$ is optimal.\hfill
$\Box$

\subsubsection{Proof of Theorem \ref{Thm:LimitPoint_Optimality}}
{\bf Proof.}
We first establish the following recursion \eqref{Eq:Var1ofPG-EXTRAPush} of PG-ExtraPush, i.e.,
\begin{align}
\label{Eq:Var1ofPG-EXTRAPush}
\left\{
\begin{array}{l}
{\bar A}{\bf z}^{t+1} = {\bar A}{\bf z}^t - \alpha \tilde{\nabla} {\bf r}({\bf x}^{t+1}) -\alpha \nabla {\bf s}({\bf x}^t) - {\bf y}^{t+1},\\
{\bf y}^{t+1} = {\bf y}^t + (\bar{A}-A){\bf z}^{t+1},\\
{\bf w}^{t+1} = A{\bf w}^{t},\\
{\bf x}^{t+1} = \diagnal({\bf w}^{t+1})^{-1}{\bf z}^{t+1},
\end{array}%
\right.
\end{align}
and then prove this theorem via exploiting \eqref{Eq:Var1ofPG-EXTRAPush}.

\textbf{1) establishing \eqref{Eq:Var1ofPG-EXTRAPush}:} By the definitions of ${\bf r}^{t+1}$ and $\mathrm{Prox}_{\alpha {\bf r}^{t+1}}$ and the ${\bf x}$-update in \eqref{Eq:PG-ExtraPush_matrixform}, it follows
\begin{align}
\label{Eq:zt+1/2-zt+1}
{\bf z}^{t+1/2} = {\bf z}^{t+1}+\alpha \tilde{\nabla}{\bf r}({\bf x}^{t+1}), \forall t \in \mathbb{N}.
\end{align}
Then the first update of \eqref{Eq:PG-ExtraPush_matrixform} implies
\begin{align}
\label{Eq:z-update}
{\bf z}^{t+1}
= {\bar A}{\bf z}^t + {\bar A}({\bf z}^t - {\bf z}^{t-1}) - \alpha (\nabla {\bf s}({\bf x}^t)-\nabla {\bf s}({\bf x}^{t-1}))
- \alpha (\tilde{\nabla} {\bf r}({\bf x}^{t+1})-\tilde{\nabla} {\bf r}({\bf x}^{t})),
\end{align}
for $t=1,2,\ldots.$
Moreover, observe that
\[
{\bf z}^1
= A{\bf z}^0 - \alpha \nabla {\bf s}({\bf x}^0) - \alpha \tilde{\nabla} {\bf r}({\bf x}^1).
\]
Summing these subgradient recursions over times 1 through $t+1$, we get
\begin{align*}
{\bf z}^{t+1} = {\bar A}{\bf z}^t + \sum_{k=0}^t (A-{\bar A}){\bf z}^k - \alpha \nabla {\bf s}({\bf x}^t) - \alpha \tilde{\nabla} {\bf r}({\bf x}^{t+1}).
\end{align*}
Furthermore, adding $(A-{\bar A}){\bf z}^{t+1}$ into both sides of the above equation and noting $A+{\bf I}_n = 2{\bar A}$, we get
\begin{align}
\label{Eq:z-update2}
{\bar A}{\bf z}^{t+1} = {\bar A}{\bf z}^t - \alpha \tilde{\nabla} {\bf r}({\bf x}^{t+1}) - \alpha \nabla {\bf s}({\bf x}^t) - {\bf y}^{t+1}.
\end{align}
Thus, based on \eqref{Eq:PG-ExtraPush_matrixform} and \eqref{Eq:z-update2}, we have \eqref{Eq:Var1ofPG-EXTRAPush}.

\textbf{2) proving subsequence convergence:}
By Property {\ref{Prop:A}},  $\{{\bf w}^t\}$ is bounded. By the last update of \eqref{Eq:Var1ofPG-EXTRAPush} and the boundedness of both $\{{\bf z}^t\}$ and $\{{\bf w}^t\}$, $\{{\bf x}^t\}$ is bounded. By the first update of \eqref{Eq:Var1ofPG-EXTRAPush} and the boundedness of $\{{\bf z}^t\}$, $\{{\bf x}^t\}$ and $\{\tilde{\nabla} {\bf r}({\bf x}^t)\}$, $\{{\bf y}^t\}$ is bounded.
Hence, there exists a convergent subsequence $\{({\bf z}, {\bf y}, {\bf w}, {\bf x})^{t_j}\}_{j=1}^{\infty}$.
Let $({\bf z}^*, {\bf y}^*, {\bf w}^*, {\bf x}^*)$ be its limit.
By {\eqref{Eq:LimofAt}}, we know that ${\bf w}^* = n \phi$ and thus that ${\bf x}^* = D_{\infty}^{-1} {\bf z}^*.$
Letting $t\to\infty$ in the second equation of (\ref{Eq:Var1ofPG-EXTRAPush}) gives ${\bf z}^* = A{\bf z}^*$, or equivalently ${\bf z}^* \in \nullspc({\bf I}_n - A)$.
Similarly, letting  $t\to\infty$ in the first equation of (\ref{Eq:Var1ofPG-EXTRAPush}) yields ${\bf y}^* + \alpha (\nabla {\bf s}({\bf x}^*)+\tilde{\nabla} {\bf r}({\bf x}^*)) = 0.$
Moreover, from the definition of ${\bf y}^t$ and the facts that both $A$ and ${\bar{A}}$ are column stochastic, it follows that ${\bf 1}_n^T {\bf y}^* =0$ and ${\bf 1}_n^T (\nabla {\bf s}({\bf x}^*)+\tilde{\nabla} {\bf r}({\bf x}^*)) =0$.
Therefore, $({\bf z}^*, {\bf y}^*, {\bf x}^*)$ satisfies the optimality conditions (\ref{Eq:1stOrderOpt}).\hfill
$\Box$

\subsubsection{Proof of Theorem \ref{Thm:ConvThm}}
The sketch of the proof is as follows: we first establish the inequality \eqref{Eq:ConvThm} holds for some fixed iteration $t$ under the bounded assumption of ${\bf v}^t$, and then prove that the inequality \eqref{Eq:ConvThm} and the boundedness of ${\bf v}^t$ hold for any $t\in \mathbb{N}$ via an inductive way, and latter give the linear convergence rate based on \eqref{Eq:ConvThm} via a recursive way.

To prove Theorem \ref{Thm:ConvThm}, we need the following lemmas.

\begin{Lemm}
\label{Lemm:optimalrelation}
For any  $({\bf z}^*, {\bf y}^*, {\bf x}^*) \in {\cal L}^*$, let ${\bf u}^*$ satisfy $(\bar{A}-A){\bf u}^* = {\bf y}^*$. Then there hold
\begin{align}
& M{\bf z}^* = {\bf 0}, \label{Eq:z*1},\\
& M^T {\bf z}^* = {\bf 0}, \label{Eq:z*2}, \\
& S{\bf v}^* + \alpha {\bf e}^*=0, \label{Eq:optcond-v*}
\end{align}
where ${\bf e}^* \triangleq
\left(\begin{array}{c}
D_{\infty}^{-1}(\tilde{\nabla} {\bf r}({\bf x}^*) + \nabla {\bf f}({\bf x}^*))\\
0
\end{array}
\right).
$
\end{Lemm}

The proof of this lemma is similar to that of \cite[Lemma 1]{Zeng-ExtraPush2016}. Thus, we omit it here.

\begin{Lemm}
\label{Lemm:IterativeRelation}
For any $t\in \mathbb{N}$, it holds
\begin{equation}
\label{Eq:IterativeRelation}
N({\bf z}^{t+1} - {\bf z}^t) = - M({\bf u}^{t+1} - {\bf u}^*) - \alpha D_{\infty}^{-1}(\tilde{\nabla}{\bf r}({\bf x}^{t+1})+\nabla {\bf s}({\bf x}^t) - \tilde{\nabla}{\bf r}({\bf x}^{*})-\nabla {\bf s}({\bf x}^*)).
\end{equation}
\end{Lemm}

This lemma follows from {\eqref{Eq:PG-ExtraPush_var2}} and the fact $M{\bf u}^* + \alpha D_{\infty}^{-1}(\tilde{\nabla}{\bf r}({\bf x}^{*})+\nabla {\bf s}({\bf x}^*))= 0$  in Theorem \ref{Thm:1stOrderOptCond}. In the following lemma, we will claim that ${\bf z}^{t+1}$ is bounded if ${\bf z}^{t-1}$ and ${\bf z}^{t}$ are bounded.

\begin{Lemm}
\label{Lemm:boundofthreesteps}
Let Assumption \ref{Assump:conectivityofgraph} and Assumption \ref{Assump:ObjFun}(i) and (iii) hold. If $\|{\bf z}^{t-1}\| \leq B$ and $\|{\bf z}^{t}\| \leq B$ for some constant $B_r\leq B<\infty$, and some $t\in \mathbb{N}_+(\triangleq \mathbb{N}\setminus \{0\})$, then $\|{\bf z}^{t+1}\|\leq C_1 B$, where $C_1 \triangleq 3+2\alpha({d^-}L_s+1),$ and $B_r$ is specified in \eqref{Eq:boundedsubgrad}.
\end{Lemm}
{\bf Proof.}
According to \eqref{Eq:z-update0} and by Property \ref{Prop:A}, it follows
\begin{align*}
\|{\bf z}^{t+1}\|
\leq 2\|{\bf z}^t\| + \|{\bf z}^{t-1}\| +\alpha L_s \|{\bf x}^t - {\bf x}^{t-1}\|+2\alpha B_r
\leq (3+2\alpha {d^-}L_s +2\alpha)B.
\end{align*}
$\Box$

The following lemma presents some basic relations that will be frequently used in the latter analysis.
\begin{Lemm}
\label{Lemm:xt-zt}
Let Assumptions \ref{Assump:conectivityofgraph} and \ref{Assump:ExisOptSolution} hold. If $\|{\bf z}^{t}\| \leq B$ for some constant $B_r\leq B<\infty$ and some $t\in \mathbb{N}$,  then
\begin{enumerate}
\item[(i)]
$\|{\bf x}^{t+1}-{\bf x}^t\| \leq {d^-}\|{\bf z}^{t+1}-{\bf z}^t\| + {(d^-)^2}nCB(1+\gamma)\gamma^t$;

\item[(ii)]
$\|{\bf x}^{t+1}-{\bf x}^*\| \leq {d^-}\|{\bf z}^{t+1}-{\bf z}^*\| + {d^-}nC\|{\bf x}^*\|\gamma^{t+1}$;

\item[(iii)]
$\|{\bf z}^{t+1}-{\bf z}^*\| \leq {d^+}\|{\bf x}^{t+1}-{\bf x}^*\| + nC\gamma\|{\bf x}^*\|\gamma^t$;

\end{enumerate}
\end{Lemm}

{\bf Proof.}
(i) Note that
\begin{align}
\label{Eq:Lemm:xt-zt-1}
\|{\bf x}^{t+1}-{\bf x}^t\|
&= \|(D^{t+1})^{-1}(D^{t+1})({\bf x}^{t+1}-{\bf x}^t)\| \nonumber\\
&\leq \|(D^{t+1})^{-1}\| \cdot \|{\bf z}^{t+1}-{\bf z}^t + (D^t-D^{t+1})(D^t)^{-1} D^t {\bf x}^t\|\nonumber\\
&\leq {d^-}\|{\bf z}^{t+1}-{\bf z}^t\| + ({d^-})^2(\|D^t-D_{\infty}\|+\|D_{\infty}-D^{t+1}\|)\|{\bf z}^t\| \nonumber\\
&\leq {d^-}\|{\bf z}^{t+1}-{\bf z}^t\| + {(d^-)}^2nCB(1+\gamma)\gamma^t,
\end{align}
where the last inequality holds for \eqref{Eq:rate-D-Dt}.
Similar to the proof of (i), we can easily prove (ii). Next, we prove (iii). Notice that
\begin{align}
\label{Eq:Lemm:xt-zt-3}
\|{\bf z}^{t+1}-{\bf z}^*\|
&= \|{\bf z}^{t+1} - D^{t+1}{\bf x}^* + D^{t+1}{\bf x}^* - {\bf z}^*\| \nonumber\\
&\leq {d^+}\|{\bf x}^{t+1}-{\bf x}^*\| + \|D^{t+1}-D_{\infty}\|\|{\bf x}^*\| \nonumber\\
&\leq {d^+}\|{\bf x}^{t+1}-{\bf x}^*\| + nC\gamma\|{\bf x}^*\|\gamma^t.
\end{align}
Thus, we end the proof.
$\Box$

As shown in Lemma \ref{Lemm:boundofthreesteps}, it requires that $\|{\bf z}^{t-1}\|$ and $\|{\bf z}^t\|$ are bounded by the same constant $B$. The following lemma gives a specific representation of $B$ under the boundedness of $\|{\bf v}^t - {\bf v}^*\|_G^2$.
\begin{Lemm}
\label{Lemm:B}
Let Assumptions \ref{Assump:conectivityofgraph}, \ref{Assump:ExisOptSolution} and \ref{Assump:WeightingMatrices} hold. If $\|{\bf v}^t - {\bf v}^*\|_G^2$ is bounded by some constant ${\cal B}$ for some $t$, i.e., $\|{\bf v}^t - {\bf v}^*\|_G^2 \leq {\cal B}$, then $\|{\bf z}^t\|$ is bounded by some constant $B$ specified as follows
\begin{align}
\label{Eq:B}
\|{\bf z}^t\| \leq B \triangleq \max\left\{\sqrt{\frac{{\cal B}}{\lambda_{\min}(\frac{N+N^T}{2})}}+\|{\bf z}^*\|,B_r\right\}.
\end{align}
\end{Lemm}

{\bf Proof.}
By the definitions of matrix $G$ and sequence $\{{\bf v}^t\}$, it is obvious that
\begin{align*}
\|{\bf z}^t - {\bf z}^*\|_{\frac{N+N^T}{2}}^2 \leq \|{\bf v}^t - {\bf v}^*\|_G^2 \leq {\cal B},
\end{align*}
which implies
\begin{align*}
\|{\bf z}^t - {\bf z}^*\| \leq \sqrt{\frac{{\cal B}}{\lambda_{\min}(\frac{N+N^T}{2})}}.
\end{align*}
Thus, we can easily claim \eqref{Eq:B}.
$\Box$

To establish the key inequality \eqref{Eq:ConvThm}, we need to develop an important inequality under the boundedness of $\|{\bf v}^{t}-{\bf v}^*\|$ as shown in the following lemma.
\begin{Lemm}
\label{Lemm:ImportIneq}
Let Assumptions \ref{Assump:conectivityofgraph}-\ref{Assump:WeightingMatrices} hold. Let $\{{\bf v}^t\}$ be a sequence generated by the iteration \eqref{Eq:PG-ExtraPush_var3} and ${\bf v}^*$ be defined in \eqref{Eq:MetricForm}. If $\|{\bf v}^{t-1}-{\bf v}^*\| \leq {\cal B}$ and $\|{\bf v}^{t}-{\bf v}^*\| \leq {\cal B}$ for some constant ${\cal B}$, and some $t\in \mathbb{N}_+$, then the following holds
\begin{align}
\label{Eq:ImportIneq}
\|{\bf v}^{t+1}-{\bf v}^*\|_G^2 - \|{\bf v}^{t}-{\bf v}^*\|_G^2
&\leq -\|{\bf v}^{t+1}-{\bf v}^t\|_G^2 + \|{\bf z}^{t+1}-{\bf z}^*\|_P^2 + \|{\bf z}^{t+1}-{\bf z}^t\|_Q^2 \nonumber\\
&+ \|{\bf u}^{t+1}-{\bf u}^*\|_R^2 +\alpha C_2 \gamma^{2t} + \alpha C_3\gamma^t.
\end{align}
where $P\triangleq \left[\frac{1}{\sigma}+\alpha\left(\frac{\eta{d_{\infty}^-}{d^-}L_s}{2}-\frac{\mu_s}{2(d^+)^2}\right)\right]{\bf I}_n$, $Q\triangleq \frac{\sigma}{2}NN^T+\alpha\frac{{d_{\infty}^-}{d^-}L_s}{2\eta}{\bf I}_n$, $R\triangleq \frac{\sigma}{2}MM^T$, $C_2 \triangleq \mu_s \left(\frac{nC\gamma \|{\bf x}^*\|}{d^+}\right)^2$, $C_3 \triangleq {d_{\infty}^-}{d^-}nCB\left[{d^-}L_s (1+\gamma)(C_1B+\|{\bf z}^*\|)+2C_1B\gamma\right]$, $B$ is specified in \eqref{Eq:B}, $\sigma>0$ and $\eta>0$ are two tunable parameters.
\end{Lemm}
{\bf Proof.}
Note that
\begin{align}
\label{Eq:SquareEqForm}
 \|{\bf v}^{t+1} - {\bf v}^*\|_G^2 - \|{\bf v}^t - {\bf v}^*\|^2_G
&= -\|{\bf v}^{t+1} - {\bf v}^t\|^2_G + \langle {\bf v}^*- {\bf v}^{t+1}, G({\bf v}^{t}-{\bf v}^{t+1}) \rangle\nonumber\\
& \ \ \   + \langle {\bf v}^*- {\bf v}^{t+1}, G^T({\bf v}^{t}-{\bf v}^{t+1}) \rangle.
\end{align}
In the following, we analyze the two inner-product terms:
\begin{align}
\label{Eq:1stTerm}
\langle {\bf v}^*- {\bf v}^{t+1}, G({\bf v}^{t}-{\bf v}^{t+1}) \rangle
& = \langle {\bf z}^*- {\bf z}^{t+1}, N^T({\bf z}^{t}-{\bf z}^{t+1}) \rangle + \langle M^T({\bf u}^*- {\bf u}^{t+1}), {\bf u}^{t}-{\bf u}^{t+1} \rangle \nonumber\\
(\because {\eqref{Eq:z*1}}, M{\bf z}^* = {\bf 0})
& = \langle {\bf z}^*- {\bf z}^{t+1}, N^T({\bf z}^{t}-{\bf z}^{t+1}) \rangle + \langle M^T({\bf u}^*- {\bf u}^{t+1}), {\bf z}^{*}-{\bf z}^{t+1} \rangle \nonumber\\
& \leq \frac{\sigma}{2}\|{\bf z}^{t}-{\bf z}^{t+1}\|_{NN^T}^2 + \frac{1}{\sigma}\|{\bf z}^{*}-{\bf z}^{t+1}\|^2 + \frac{\sigma}{2}\|{\bf u}^{*}-{\bf u}^{t+1}\|_{MM^T}^2,
\end{align}
where $\sigma>0$ is a tunable parameter, and
\begin{align}
\label{Eq:2ndTerm0}
\langle {\bf v}^*- {\bf v}^{t+1}, G^T({\bf v}^{t}-{\bf v}^{t+1}) \rangle
& = \langle {\bf v}^*- {\bf v}^{t+1}, S{\bf v}^{t+1}+\alpha {\bf e}^t \rangle  \ (\because {\eqref{Eq:MetricForm}})\nonumber\\
& = \langle {\bf v}^*- {\bf v}^{t+1}, S({\bf v}^{t+1}-{\bf v}^{*})+\alpha ({\bf e}^t-{\bf e}^*) \rangle \ (\because {\eqref{Eq:optcond-v*}}) \nonumber\\
(\because S = -S^T) \ & = \alpha \langle {\bf v}^*- {\bf v}^{t+1}, {\bf e}^t-{\bf e}^*\rangle \nonumber\\
& = \alpha \langle {\bf z}^*-{\bf z}^{t+1}, D_{\infty}^{-1}(\tilde{\nabla}{\bf r}({\bf x}^{t+1})-\tilde{\nabla}{\bf r}({\bf x}^*) + \nabla {\bf s}({\bf x}^t) -\nabla {\bf s}({\bf x}^*))\rangle \nonumber\\
&= \alpha \langle D_{\infty}^{-1}({\bf z}^*-{\bf z}^{t+1}), \tilde{\nabla}{\bf r}({\bf x}^{t+1})-\tilde{\nabla}{\bf r}({\bf x}^*) \rangle \nonumber\\
&+ \alpha \langle D_{\infty}^{-1}({\bf z}^*-{\bf z}^{t+1}), \nabla {\bf s}({\bf x}^t) -\nabla {\bf s}({\bf x}^*) \rangle \quad (\triangleq \alpha(T_1 + T_2)).
\end{align}
Next, we give upper bounds of $T_1$ and $T_2$, respectively.
\begin{align}
\label{Eq:T1}
T_1
&= \langle {\bf x}^* - {\bf x}^{t+1}, \tilde{\nabla}{\bf r}({\bf x}^{t+1})-\tilde{\nabla}{\bf r}({\bf x}^*)\rangle + \langle ((D^{t+1})^{-1}-D_{\infty}^{-1}){\bf z}^{t+1}, \tilde{\nabla}{\bf r}({\bf x}^{t+1})-\tilde{\nabla}{\bf r}({\bf x}^*)\rangle \nonumber\\
&\leq  \langle ((D^{t+1})^{-1}-D_{\infty}^{-1}){\bf z}^{t+1}, \tilde{\nabla}{\bf r}({\bf x}^{t+1})-\tilde{\nabla}{\bf r}({\bf x}^*)\rangle \quad (\because {\bf r} \ \text{is convex})\nonumber\\
&\leq 2{d^-}{d_{\infty}^-}nCC_1B^2 \gamma^{t+1} \quad (\because \eqref{Eq:rate-invD-invDt}, \text{Assumption \ref{Assump:ObjFun}(iii)}, \text{Lemma \ref{Lemm:boundofthreesteps}}),
\end{align}
and
\begin{align}
\label{Eq:T20}
T_2
&= \langle D_{\infty}^{-1}({\bf z}^* - {\bf z}^{t+1}), \nabla {\bf s}({\bf x}^{t+1})-\nabla {\bf s}({\bf x}^*) \rangle + \langle D_{\infty}^{-1}({\bf z}^* - {\bf z}^{t+1}), \nabla {\bf s}({\bf x}^{t})-\nabla {\bf s}({\bf x}^{t+1}) \rangle\nonumber\\
&\leq -\mu_s \|{\bf x}^{t+1}-{\bf x}^*\|^2 + {d_{\infty}^-}\|{\bf z}^{t+1}-{\bf z}^*\|\cdot L_s\|{\bf x}^t - {\bf x}^{t+1}\| \quad (\because \eqref{Eq:StrongCVX}, \eqref{Eq:def-d}, \eqref{Eq:LipC-Grad}).
\end{align}
By Lemma \ref{Lemm:xt-zt}(iii), it follows
\begin{align*}
\|{\bf z}^{t+1}-{\bf z}^*\|^2 \leq 2(d^+)^2 \|{\bf x}^{t+1}-{\bf x}^*\|^2 + 2(nC\gamma \|{\bf x}^*\|)^2 \gamma^{2t},
\end{align*}
which implies
\begin{align}
\label{Eq:xt+1-xt}
\|{\bf x}^{t+1}-{\bf x}^*\|^2 \geq \frac{1}{2(d^+)^2} \|{\bf z}^{t+1}-{\bf z}^*\|^2 - \left(\frac{nC\gamma \|{\bf x}^*\|}{d^+}\right)^2 \gamma^{2t}.
\end{align}
By Lemma \ref{Lemm:xt-zt}(i), it shows
\begin{align}
\label{Eq:timesoftwoterm}
&\|{\bf z}^{t+1}-{\bf z}^*\|\|{\bf x}^t - {\bf x}^{t+1}\| \nonumber\\
&\leq {d^-}\|{\bf z}^{t+1}-{\bf z}^*\|  \|{\bf z}^{t+1} - {\bf z}^t\| + {(d^-)^2}nCB(1+\gamma)\gamma^t\|{\bf z}^{t+1}-{\bf z}^*\| \nonumber\\
&\leq \frac{{d^-}}{2}(\eta\|{\bf z}^{t+1}-{\bf z}^*\|^2 + \eta^{-1}\|{\bf z}^{t+1} - {\bf z}^t\|^2) + {(d^-)^2}nCB(1+\gamma)(C_1B+\|{\bf z}^*\|)\gamma^t,
\end{align}
where $\eta>0$ is a tunable parameter.
Substituting {\eqref{Eq:xt+1-xt}} and \eqref{Eq:timesoftwoterm} into {\eqref{Eq:T20}}, then we have
\begin{align}
\label{Eq:T2}
T_2
&\leq \left(\frac{\eta{d_{\infty}^-}{d^-}L_s}{2}-\frac{\mu_s}{2(d^+)^2}\right)\|{\bf z}^{t+1}-{\bf z}^*\|^2 + \frac{{d_{\infty}^-}{d^-}L_s}{2\eta}\|{\bf z}^{t+1}-{\bf z}^t\|^2 \nonumber\\
&+\mu_s \left(\frac{nC\gamma \|{\bf x}^*\|}{d^+}\right)^2 \gamma^{2t} + {d_{\infty}^-}{(d^-)^2}L_s nCB(1+\gamma)(C_1B+\|{\bf z}^*\|)\gamma^t.
\end{align}
Plugging \eqref{Eq:T1} and \eqref{Eq:T2} into \eqref{Eq:2ndTerm0}, it becomes
\begin{align}
\label{Eq:2ndTerm}
&\langle {\bf v}^*- {\bf v}^{t+1}, G^T({\bf v}^{t}-{\bf v}^{t+1}) \rangle \nonumber\\
&\leq \alpha\left(\frac{\eta{d_{\infty}^-}{d^-}L_s}{2}-\frac{\mu_s}{2(d^+)^2}\right)\|{\bf z}^{t+1}-{\bf z}^*\|^2 + \alpha\frac{{d_{\infty}^-}{d^-}L_s}{2\eta}\|{\bf z}^{t+1}-{\bf z}^t\|^2 \nonumber\\
&+\alpha\mu_s \left(\frac{nC\gamma \|{\bf x}^*\|}{d^+}\right)^2 \gamma^{2t} + \alpha
{d_{\infty}^-}{d^-}nCB\left[{d^-}L_s (1+\gamma)(C_1B+\|{\bf z}^*\|)+2C_1B\gamma\right]\gamma^t.
\end{align}
Substituting \eqref{Eq:1stTerm} and \eqref{Eq:2ndTerm} into \eqref{Eq:SquareEqForm}, we can conclude \eqref{Eq:ImportIneq}.
$\Box$

Based on Lemma \ref{Lemm:ImportIneq}, we can establish \eqref{Eq:ConvThm} for some $\delta>0$ and some $t\in \mathbb{N}$ under some assumptions as shown in the following lemma.

\begin{Lemm}
\label{Lemm:estab-ImportIneq}
Under conditions of Theorem 1, if $\|{\bf v}^{t-1}-{\bf v}^*\| \leq {\cal B}$ and $\|{\bf v}^{t}-{\bf v}^*\| \leq {\cal B}$ for some constant ${\cal B}$ and some $t\in \mathbb{N}_+$, then the inequality \eqref{Eq:ConvThm} holds for some constants $\delta$ and $\Gamma_0$.
\end{Lemm}

{\bf Proof.}
In order to establish \eqref{Eq:ConvThm} for some $\delta>0$ and $\Gamma_0$, in light of Lemma \ref{Lemm:ImportIneq}, it is sufficient to show that the right-hand side of \eqref{Eq:ImportIneq} is no more than $-\delta \|{\bf v}^{t+1}-{\bf v}^*\|_G^2 +\Gamma_0 \gamma^t$, which implies
\begin{align}
&\|{\bf z}^{t+1}-{\bf z}^*\|^2_{P_1} + \|{\bf z}^{t+1}-{\bf z}^t\|_{Q_1}^2 \geq \|{\bf u}^{t+1}-{\bf u}^*\|^2_{R_1}, \label{Eq:matrixpart}\\
&\alpha C_2 \gamma^{2t} +\alpha C_3 \gamma^t \leq \Gamma_0 \gamma^t, \label{Eq:exponentpart}
\end{align}
where
$
P_1 = (\alpha \bar{\mu} - \frac{\alpha \bar{\eta}}{2} -\frac{1}{\sigma}){\bf I}_n - \delta \frac{N+N^T}{2},
$
$
Q_1 = \frac{N^T+N}{2} - \frac{\sigma}{2}NN^T - \frac{\alpha \bar{L}^2}{2\eta}{\bf I}_n
$
and
$
R_1 = \frac{\sigma}{2}MM^T + \delta {(\frac{M+M^T}{2})},
$
$\bar{\mu} = \frac{\mu_s}{(2d^+)^2}$, $\bar{\eta} = {d_{\infty}^-}{d^-}L_s \eta$ and $\bar{L} = {d_{\infty}^-}{d^-}L_s$.

Let
$c_4 \triangleq (\bar{\mu}- \frac{\bar{\eta}}{2})+\sqrt{\Delta_1}$,
$c_5 \triangleq \frac{{\bar L}^2}{\bar{\eta}}$,
$c_{6} \triangleq \frac{2c_4c_5 + 12c_1{\bar L}^2}{c_4^2}$,
$c_{7} \triangleq \frac{\lambda_{\min}^2(N^T+N)}{4c_3}$,
$c_{8} \triangleq a(c_{7}+2)-(2-c_{7})$ for some positive constant $a\in (0,1)$,
$\Delta_3 \triangleq \lambda_{\min}^2(N^T+N) - 4c_3c_{6}$.
According to the similar proof of \cite[Theorem 4]{Zeng-ExtraPush2016}, we can claim that if the following conditions hold
\begin{align}
\label{Eq:Cond-a}
\frac{2-c_{7}}{2+c_{7}}<a<1,
\end{align}
\begin{align}
\label{Eq:CondonFun}
{\bar \mu}
> \big(\sqrt{\frac{6c_1}{1-a^2}} + \frac{1}{c_8}\sqrt{\frac{1-a^2}{6c_1}}\big) \bar{L},
\end{align}
\begin{align}
\label{Eq:Condoneta}
{\bar \mu}\left(1-\sqrt{1-\frac{4{\bar L}^2}{c_8{\bar \mu}^2}}\right) < \bar{\eta} < \min \left\{{\bar \mu}\big(1+\sqrt{1-\frac{4{\bar L}^2}{c_8{\bar \mu}^2}}\big), 2(\bar{\mu} - \sqrt{\frac{6c_1}{1-a^2}}\bar{L})\right\},
\end{align}
\begin{align}
\label{Eq:Condonsigma}
\frac{\lambda_{\min}(N^T+N) - \sqrt{\Delta_3}}{2c_3}< \sigma <\frac{\lambda_{\min}(N^T+N) + \sqrt{\Delta_3}}{2c_3},
\end{align}
\begin{align}
\label{Eq:Condonalpha}
\frac{{\bar \mu} - \frac{\bar{\eta}}{2} - \sqrt{\Delta_1}}{3c_1{\bar L}^2\sigma} < \alpha < \min\big\{\frac{{\bar \mu} - \frac{\bar{\eta}}{2} + \sqrt{\Delta_1}}{3c_1{\bar L}^2\sigma}, \frac{-\frac{{\bar L}^2}{2\bar{\eta}}+ \sqrt{\Delta_2}}{3c_1{\bar L}^2 \sigma}\big\},
\end{align}
then \eqref{Eq:matrixpart} holds for some positive constant $\delta$ as specified in \eqref{Eq:Condondelta}.

Taking
\begin{align}
\label{Eq:def-gamma0}
\Gamma_0 = \alpha (C_2 + C_3),
\end{align}
where $C_2$ and $C_3$ are specified in Lemma \ref{Lemm:ImportIneq}, we can easily establish the inequality \eqref{Eq:exponentpart}.
Thus, the proof of this lemma is completed.
$\Box$

According to Lemma \ref{Lemm:estab-ImportIneq}, the key inequality \eqref{Eq:ConvThm} holds for some fixed iteration $t$ if ${\bf v}^t$ is bounded. In the following lemma, we will show that when $t$ is sufficiently large, ${\bf v}^{t+1}$ is also bounded if ${\bf v}^t$ is bounded and the relation \eqref{Eq:ConvThm} holds at the $t$-th iteration.
\begin{Lemm}
\label{Lemm:T*}
Let Assumptions \ref{Assump:conectivityofgraph}-\ref{Assump:WeightingMatrices} hold. If at the $t$-th iteration, $\|{\bf v}^t - {\bf v}^*\| \leq {\cal B}$ for some constant ${\cal B}$, and the relation $\|{\bf v}^t - {\bf v}^*\|_G^2 \geq (1+\delta)\|{\bf v}^{t+1}-{\bf v}^*\|_G^2-\Gamma_0 \gamma^t$ holds for some constants $\delta, \Gamma_0>0$, then it holds
\begin{align}
\label{Eq:bounded-t+1}
\|{\bf v}^{t+1}-{\bf v}^*\|_G^2 \leq {\cal B}
\end{align}
for all $t\geq T^*$ with
\begin{align}
\label{Eq:T*}
T^* = \left\lceil\log_{\gamma} \frac{\delta {\cal B}}{\Gamma_0}\right\rceil +1,
\end{align}
where $\lceil b \rceil$ denotes the integer no less than $b$ for any $b \in \mathbb{R}$.
\end{Lemm}

{\bf Proof.}
By the definition of \eqref{Eq:T*}, it implies
\begin{align}
\label{Eq:gammat}
\Gamma_0\gamma^t \leq \delta {\cal B}
\end{align}
for any $t\geq T^*$. This together with the relation
\[
\|{\bf v}^{t+1}-{\bf v}^*\|_G^2 \leq \frac{\|{\bf v}^{t}-{\bf v}^*\|_G^2}{1+\delta} + \frac{\delta {\cal B}}{1+\delta}
\]
yield $\|{\bf v}^{t+1}-{\bf v}^*\|_G^2 \leq {\cal B}$.
$\Box$

With these lemmas, we can prove our main theorem.

{\bf (Proof for Theorem \ref{Thm:ConvThm}):}
Let ${\cal B}\triangleq \max_{0\leq t \leq T^*} \|{\bf v}^t - {\bf v}^*\|_G^2$, where $T^*$ is specified in \eqref{Eq:T*}.
By Lemma \ref{Lemm:estab-ImportIneq}, the inequality \eqref{Eq:ConvThm} holds for some fixed $t$ under the boundedness of $\|{\bf v}^t-{\bf v}^*\|$ and other conditions. In the following, we show that the inequality \eqref{Eq:ConvThm} and the boundedness of $\|{\bf v}^t-{\bf v}^*\|$ hold for any $t$.

We first prove these for the first $T^*$ iterates. By the definition of ${\cal B}$, it is obvious that $\|{\bf v}^t - {\bf v}^*\|_G^2 \leq {\cal B}$ when $t \in \{0,\ldots,T^*\}$. Moreover, by Lemma \ref{Lemm:estab-ImportIneq}, the relation \eqref{Eq:ConvThm} also holds for any $t \in \{0,\ldots,T^*\}$.

Next, we prove the inequality \eqref{Eq:ConvThm} and the boundedness of $\|{\bf v}^t-{\bf v}^*\|$ hold for any $t \geq T^*$ via an inductive way.
\begin{enumerate}
\item[(a)] \textbf{Base step:} when $t=T^*$, we have the following relations:
\begin{align}
& \|{\bf v}^{T^*-1}-{\bf v}^*\|_G^2 \leq {\cal B}, \label{Eq:basestep-T*-1}\\
& \|{\bf v}^{T^*}-{\bf v}^*\|_G^2 \leq {\cal B}, \label{Eq:basestep-T*}\\
& \|{\bf v}^{T^*}-{\bf v}^*\|_G^2 \geq (1+\delta)\|{\bf v}^{T^*+1}-{\bf v}^*\|_G^2 - \Gamma_0 \gamma^{T^*}. \label{Eq:basestep-relation}
\end{align}

\item[(b)] \textbf{Hypothesis step:} We assume that the induction hypothesis is true at the $t$-th iteration for some $t \geq T^*$, i.e.,
\begin{align}
& \|{\bf v}^{t-1}-{\bf v}^*\|_G^2 \leq {\cal B}, \label{Eq:hypostep-T*-1}\\
& \|{\bf v}^{t}-{\bf v}^*\|_G^2 \leq {\cal B}, \label{Eq:hypostep-T*}\\
& \|{\bf v}^{t}-{\bf v}^*\|_G^2 \geq (1+\delta)\|{\bf v}^{t+1}-{\bf v}^*\|_G^2 - \Gamma_0 \gamma^{t}. \label{Eq:hypostep-relation}
\end{align}

\item[(c)] \textbf{Inductive step:} We then show that the above relations hold for the $(t+1)$-th iteration.
By \eqref{Eq:hypostep-T*} and Lemma \ref{Lemm:T*}, it holds $\|{\bf v}^{t+1}-{\bf v}^*\|_G^2 \leq {\cal B}$. Based on the boundedness of $\|{\bf v}^{t+1}-{\bf v}^*\|_G^2$ and $\|{\bf v}^{t}-{\bf v}^*\|_G^2$, and by Lemma \ref{Lemm:estab-ImportIneq}, the inequality \eqref{Eq:ConvThm} holds for the $(t+1)$-th iteration, i.e.,
\[
\|{\bf v}^{t+1}-{\bf v}^*\|_G^2 \geq (1+\delta)\|{\bf v}^{t+2}-{\bf v}^*\|_G^2 - \Gamma_0 \gamma^{t}.
\]
\end{enumerate}
By induction, we conclude that these relations hold for all $t$.

In the following, we establish the linear rate of the sequence $\{{\bf x}^t\}$ based on \eqref{Eq:ConvThm}.
Let $\tau \triangleq \max\{\frac{1}{1+\delta},\gamma\}$. From \eqref{Eq:ConvThm}, for any $t$, there holds
\begin{align*}
\|{\bf v}^t - {\bf v}^*\|_G^2
&\leq \frac{1}{1+\delta}\|{\bf v}^{t-1} - {\bf v}^*\|_G^2 + \Gamma_0 \frac{\gamma^{t-1}}{1+\delta} \nonumber\\
&\leq \tau \|{\bf v}^{t-1} - {\bf v}^*\|_G^2 + \Gamma_0 \tau^t \nonumber\\
&\leq \tau^t \|{\bf v}^{0} - {\bf v}^*\|_G^2 + t\Gamma_0\tau^t.
\end{align*}
Taking a $\rho \in (\tau,1)$. Let $\xi \triangleq \frac{2}{\frac{\rho}{\tau}\ln (\frac{\rho}{\tau})}$. Then for any $t\in \mathbb{N}$, it holds
\[
\left(\frac{\rho}{\tau}\right)^t > \frac{t}{\xi}.
\]
As a consequence, we have
\begin{align}
\label{Eq:linearrate-vt}
\|{\bf v}^t - {\bf v}^*\|_G^2
&\leq \rho^t \|{\bf v}^{0} - {\bf v}^*\|_G^2 + (\Gamma_0\xi)\frac{t}{\xi}\left(\frac{\tau}{\rho}\right)^t\rho^t \nonumber\\
&\leq (\|{\bf v}^{0} - {\bf v}^*\|_G^2 + \Gamma_0\xi)\rho^t.
\end{align}
By the definitions of $G$ and sequence $\{{\bf v}^t\}$ (see, \eqref{Eq:MetricForm}), \eqref{Eq:linearrate-vt} implies
\begin{align}
\label{Eq:linearrate-zt}
\|{\bf z}^t - {\bf z}^*\|_{\frac{N+N^T}{2}}^2 \leq \|{\bf v}^t - {\bf v}^*\|_G^2 \leq (\|{\bf v}^{0} - {\bf v}^*\|_G^2 + \Gamma_0\xi)\rho^t.
\end{align}
By Assumption \ref{Assump:WeightingMatrices}, the matrix $N+N^T$ is positive definite. Thus, \eqref{Eq:linearrate-zt} implies
\begin{align}
\label{Eq:linearrate-zt1}
\|{\bf z}^t - {\bf z}^*\| \leq \sqrt{\frac{\|{\bf v}^{0} - {\bf v}^*\|_G^2 + \Gamma_0\xi}{\lambda_{\min}(\frac{N+N^T}{2})}}(\sqrt{\rho})^t.
\end{align}
Furthermore, by Lemma \ref{Lemm:xt-zt}(ii), \eqref{Eq:linearrate-zt1} implies
\begin{align*}
\|{\bf x}^t- {\bf x}^*\|
&\leq {d^-}\left(\sqrt{\frac{\|{\bf v}^{0} - {\bf v}^*\|_G^2 + \Gamma_0\xi}{\lambda_{\min}(\frac{N+N^T}{2})}}(\sqrt{\rho})^t + nC\|{\bf x}^*\|\gamma^t\right) \\
&\leq {d^-}\left(\sqrt{\frac{\|{\bf v}^{0} - {\bf v}^*\|_G^2 + \Gamma_0\xi}{\lambda_{\min}(\frac{N+N^T}{2})}} + nC\|{\bf x}^*\|\right)(\sqrt{\rho})^t,
\end{align*}
where the second inequality holds for $\gamma\leq \tau<\rho<\sqrt{\rho}<1.$
Let
\begin{align}
\label{Eq:def-gamma}
\Gamma \triangleq {d^-}\left(\sqrt{\frac{\|{\bf v}^{0} - {\bf v}^*\|_G^2 + \Gamma_0\xi}{\lambda_{\min}(\frac{N+N^T}{2})}} + nC\|{\bf x}^*\|\right),
\end{align}
then we get \eqref{Eq:LinConvThm}. Thus, we end the proof.
$\Box$

\section{Numerical Experiments}
\label{sc:experiment}

In this section, we provide a series of numerical experiments to show the effectiveness of the proposed algorithms via comparing to Subgradient-Push algorithm. In these experiments, the connected network and its corresponding mixing matrix $A$ are generated randomly.

\subsection{Decentralized Geometric Median}
Consider a decentralized geometric median problem. Each agent $i\in \{1,\cdots,n\}$ holds a vector $b_{(i)} \in \mathbb{R}^p$, and all the agents collaboratively calculate the geometric median $x\in \mathbb{R}^p$ of all $b_{(i)}$. This task can be formulated as solving the following minimization problem:
\begin{equation}
\label{Eq:DistributedGM}
x^* \gets \mathop{\mathrm{argmin}}_{x\in \mathbb{R}^p} f(x) = \sum_{i=1}^n \|x-b_{(i)}\|_2.
\end{equation}

The geometric median problem is solved by P-ExtraPush over directed networks. The proximity operator $\mathrm{prox}_{\alpha r_i}$ has an explicit solution, for any $u \in \mathbb{R}^p$,
\begin{align*}
\mathrm{prox}_{\alpha r_i}(u) = b_{(i)} - \frac{b_{(i)}-u}{\|b_{(i)}-u\|_2}\mathrm{max}\{\|b_{(i)}-u\|_2 - \alpha,0\}.
\end{align*}

\begin{figure}[!t]
\begin{minipage}[b]{0.99\linewidth}
\centering
\includegraphics*[scale=0.42]{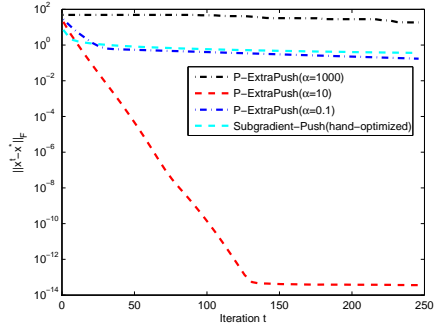}
\end{minipage}
\caption{ Experiment results for decentralized geometric median. Trends of $\|{\bf x}^t-{\bf x}^{*}\|_F$, where ${\bf x}^*$ is the limitation of ${\bf x}^t$, which is taken as the iterate at $t=1000$, i.e., ${\bf x}^{1000}$.
}
\label{Fig:DGM}
\end{figure}

We set $n=10$ and $p=256$, that is, each point $b_{(i)} \in \mathbb{R}^p.$ Data $b_{(i)}$ are generated following the i.i.d. Gaussian distribution. The algorithm starts from $z_{(i)}^0 = b_{(i)}, \forall i$. We use three different step sizes $\alpha$ to show the effect of the step size. The numerical results are reported in Fig. {\ref{Fig:DGM}}.
From Fig. \ref{Fig:DGM}, P-ExtraPush can adopt a large range of step size. More specifically, with a proper step size (say, $\alpha =10$), P-ExtraPush converges linearly and is significantly faster than Subgradient-Push algorithm even with the hand-optimized step size.

\subsection{Decentralized $\ell_1$ Regularized Least Squares Regression}

\begin{figure}[!t]
\begin{minipage}[b]{0.99\linewidth}
\centering
\includegraphics*[scale=0.42]{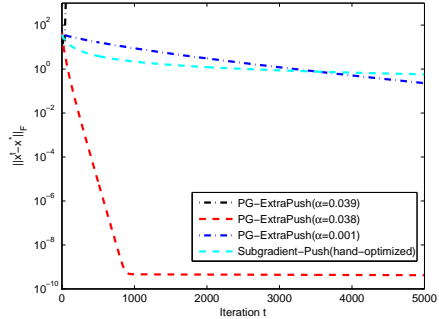}
\end{minipage}
\hfill
\caption{ Experiment results for decentralized $\ell_1$ regularized least squares regression. Trends of $\|{\bf x}^t-{\bf x}^{*}\|_F$, where ${\bf x}^*$ is the limitation of ${\bf x}^t$, which is taken as the iterate at $t=10000$.
}
\label{Fig:DL1LS}
\end{figure}

We consider the following decentralized $\ell_1$ regularized least squares regression problem, i.e.,
\begin{equation}
\label{Eq:DistributedLS}
x^* \gets \mathop{\mathrm{argmin}}_{x\in \mathbb{R}^p} f(x) = \sum_{i=1}^n f_i(x),
\end{equation}
where $f_i(x) = \frac{1}{2} \|B_{(i)} x-b_{(i)}\|_2^2+\lambda_i \|x\|_1, B_{(i)} \in \mathbb{R}^{m_i\times p}, b_{(i)} \in \mathbb{R}^{m_i}$ for $i=1, \ldots,n,$ $\|x\|_1 = \sum_{i=1}^p |x_i|$.
In this experiment, we take $n=10$, $p=256$, and $m_i = 150$ for $i=1,\ldots,n$. In this case, the proximity operator of $\ell_1$-norm is the soft shrinkage function. The experiment result is illustrated in Fig. {\ref{Fig:DL1LS}}. From Fig. {\ref{Fig:DL1LS}}, $\alpha = 0.038$ is a critical value of step size in the sense that the algorithm will diverge once $\alpha$ is bigger than this value, and with this proper step size, PG-ExtraPush converges linearly and is faster than Subgradient-Push. Moreover, a smaller step size generally implies a slower convergence rate.

\subsection{Decentralized Quadratic Programming}

We use decentralized quadratic programming as an example to show that how PG-ExtraPush solves a constrained optimization problem. Each agent $i \in \{1,\cdots,n\}$ has a local quadratic objective $\frac{1}{2} x^TQ_ix + h_i^Tx$ and a local linear constraint $a_i^T x \leq b_i,$ where the symmetric positive semidefinite matrix $Q_i \in \mathbb{R}^{p\times p}$, the vectors $h_i \in \mathbb{R}^p$ and $a_i \in \mathbb{R}^p,$ and the scalar $b_i \in \mathbb{R}$ are stored at agent $i$. The agents collaboratively minimize the average of the local objectives subject to all local constraints. The quadratic program is:
\begin{align*}
\mathrm{min}_x \sum_{i=1}^n \big(\frac{1}{2}x^TQ_ix + h_i^Tx\big), \  s.t. \ a_i^Tx \leq b_i,  i=1,\ldots,n.
\end{align*}
We recast it as
\begin{align}
\label{DistributedQP1}
\mathrm{min}_x \sum_{i=1}^n \big(\frac{1}{2}x^TQ_ix + h_i^Tx + {\cal I}(a_i^Tx-b_i)\big),
\end{align}
where
\[
{\cal I}(c) =
\left\{
\begin{array}{l}
0, \quad \mathrm{if} \ c \leq 0,\\
+\infty, \quad \mathrm{otherwise},
\end{array}%
\right.
\]
is an indicator function. Setting $s_i(x) = \frac{1}{2}x^TQ_ix + h_i^Tx$ and $r_i(x) = {\cal I}(a_i^Tx-b_i)$, it has the form of \eqref{Eq:multi-agentOPT} and can be solved by PG-ExtraPush. The proximity operator $\mathrm{prox}_{\alpha r_i}$ has an explicit solution
\begin{align*}
\mathrm{prox}_{\alpha r_i}(u) = \left\{
\begin{array}{ll}
u, &\mathrm{if} \ a_i^Tu \leq b_i \leq 0,\\
u + \frac{(b_i-a_i^Tu)a_i}{\|a_i\|_2^2}, &\mathrm{otherwise}.
\end{array}%
\right.
\end{align*}

\begin{figure}[t]
\begin{minipage}[b]{0.99\linewidth}
\centering
\includegraphics*[scale=0.42]{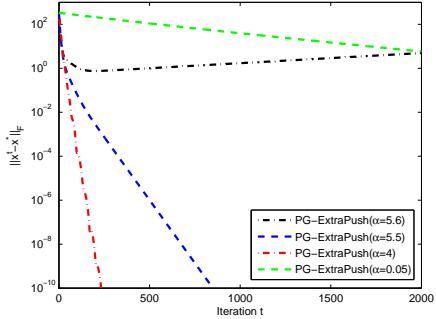}
\end{minipage}
\hfill
\caption{ Experiment results for decentralized quadratic programming with symmetric positive semidefinite $Q_i's$. Trends of $\|{\bf x}^t-{\bf x}^{*}\|_F$, where ${\bf x}^*$ is the limitation of ${\bf x}^t$, which is taken as the iterate at $t=10000$.
}
\label{Fig:DQP}
\end{figure}

Consider $n=10$ and $p=256$. For any agent $i$, $Q_i$ is a positive semidefinite symmetric matrix, $h_i$, $a_i$ and $b_i$ are generated from i.i.d. Gaussian distribution. Four different step sizes are used to show the effect of the step size. The experiment result is presented in Fig. \ref{Fig:DQP}. Since Subgradient-Push is not appropriately used to solve this problem, so we show the performance of PG-ExtraPush only without any comparison in this case.
As show in Fig. \ref{Fig:DQP}, PG-ExtraPush can also adopt a large range of the step size parameter and $\alpha = 5.5$ is a critical value in this case in the sense that PG-ExtraPush may diverge if a larger step size is adopted.
With a proper step size (say, $\alpha = 4$), PG-ExtraPush performs the similar linear convergence rate when all $Q_i$ are positive semidefinite.

\subsection{Nonconvex Decentralized $\ell_q (0\leq q<1)$  Regularization}

\begin{figure}[!t]
\begin{minipage}[b]{.33\linewidth}
\centering
\includegraphics*[scale=0.35]{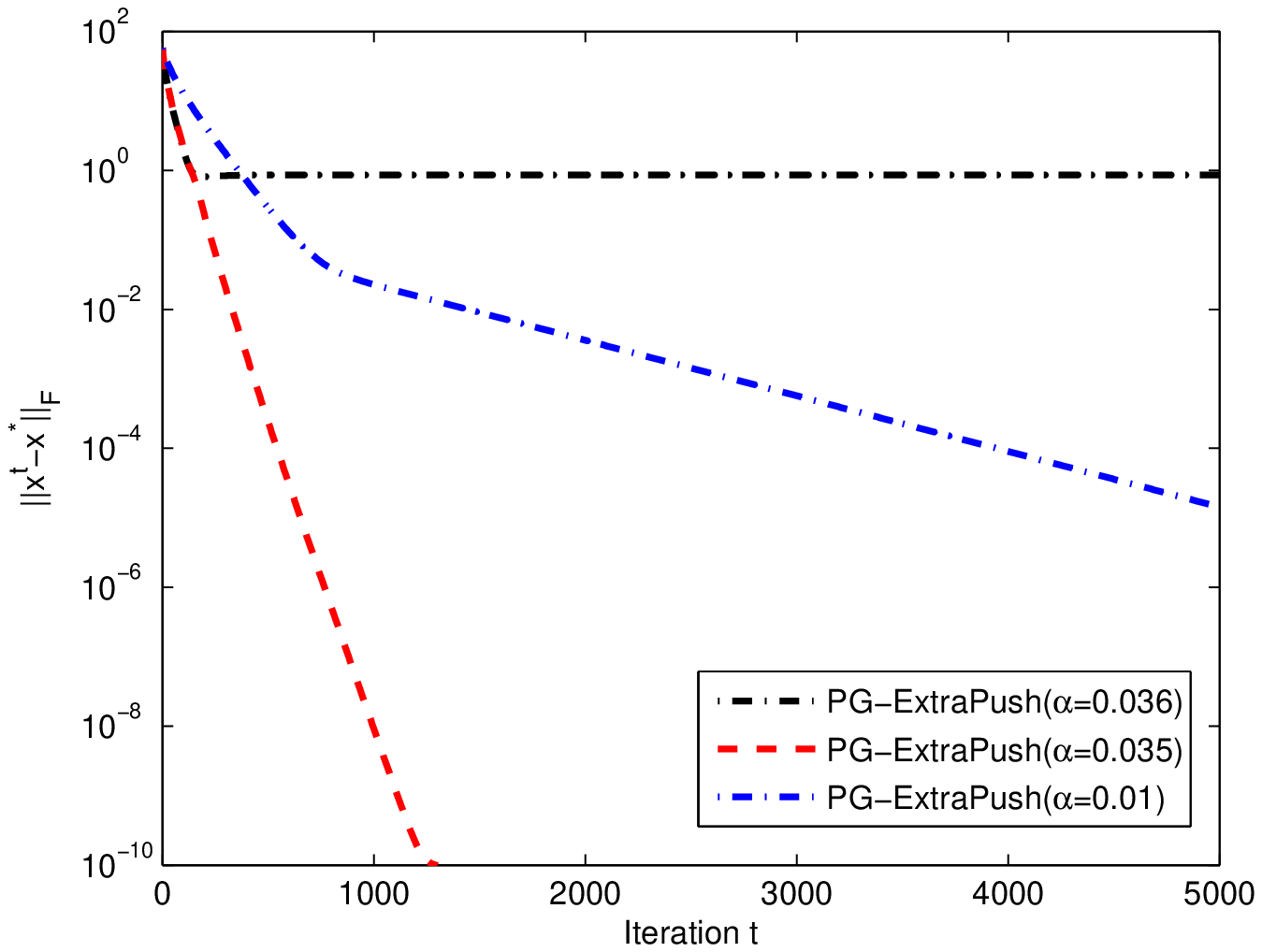}
\centerline{{\small (a) $\ell_0$}}
\end{minipage}
\begin{minipage}[b]{0.33\linewidth}
\centering
\includegraphics*[scale=0.35]{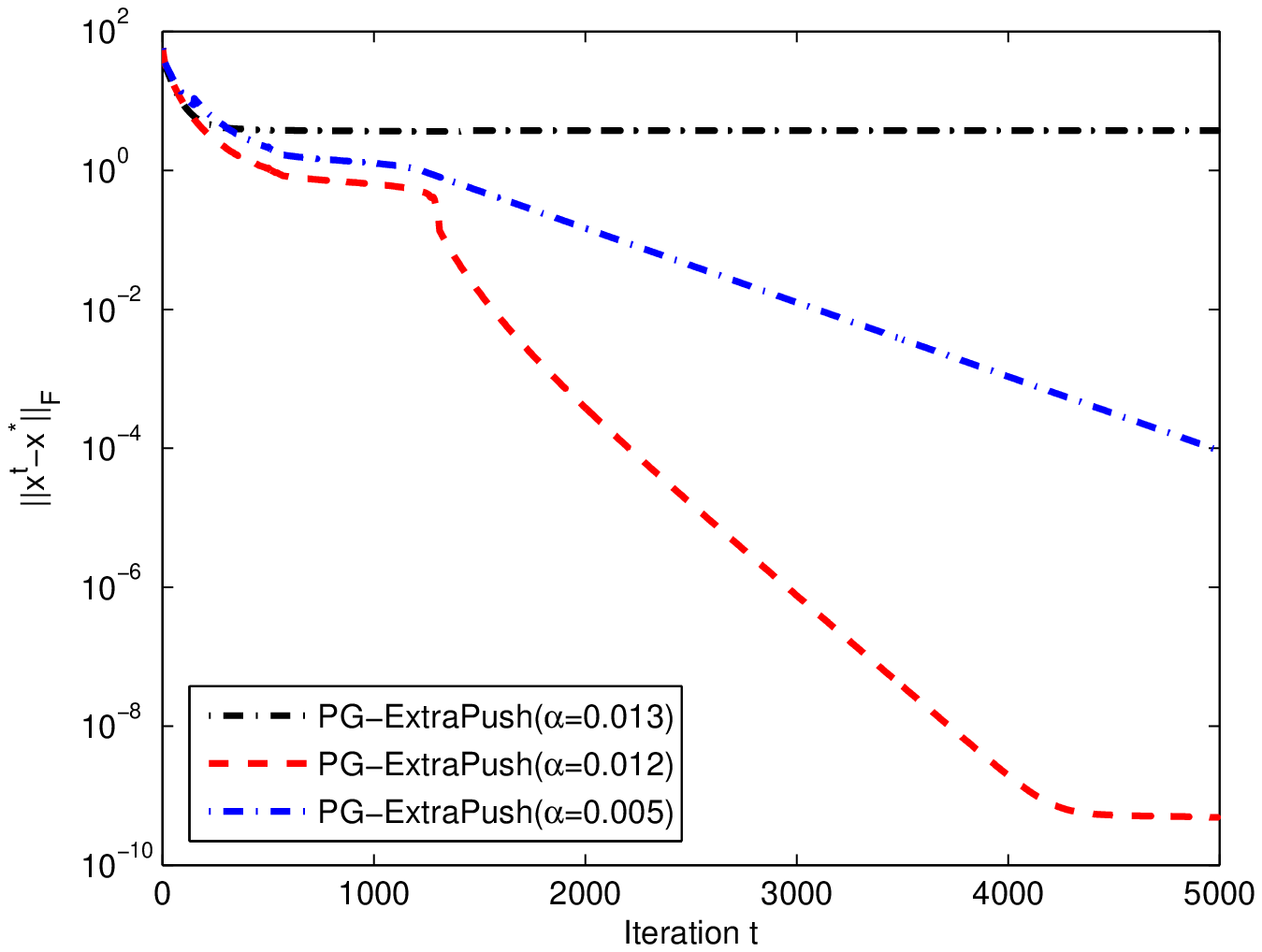}
\centerline{{\small (b) $\ell_{1/2}$}}
\end{minipage}
\hfill
\begin{minipage}[b]{0.33\linewidth}
\centering
\includegraphics*[scale=0.35]{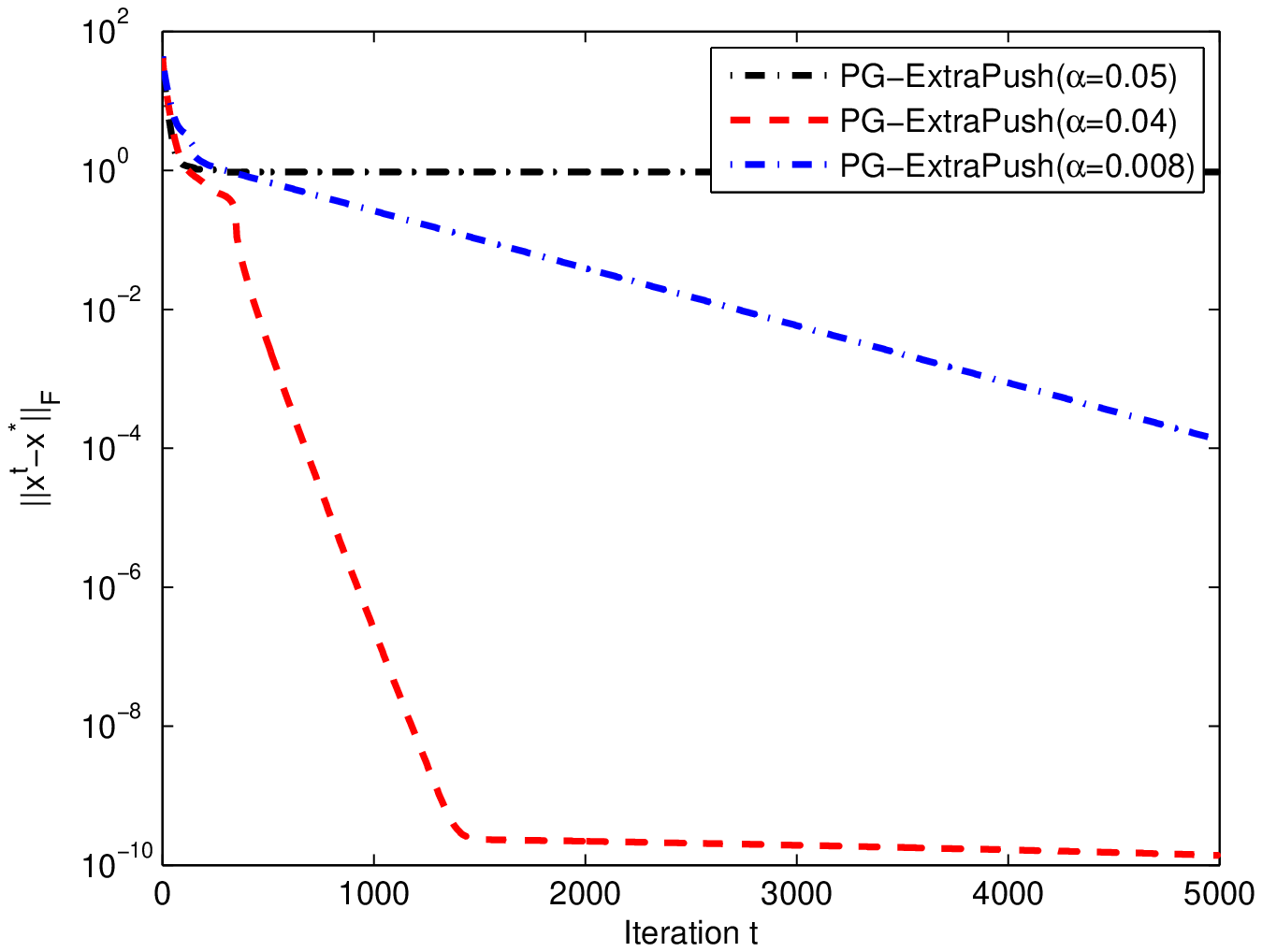}
\centerline{{\small (c) $\ell_{2/3}$}}
\end{minipage}
\caption{ Experiment results for the decentralized $\ell_q (0\leq q<1)$  regularized least squares regression. Trends of $\|{\bf x}^t-{\bf x}^{*}\|_F$, where ${\bf x}^*$ is the limitation of ${\bf x}^t$, which is taken as the iterate at $t=10000$.
}
\label{Fig:DLqLS}
\end{figure}

We apply the proposed algorithm to solve the following nonconvex decentralized $\ell_q$ $(0\leq q <1)$ regularized least squares regression problem, i.e.,
\begin{equation}
\label{Eq:DistributedLS}
x^* \gets \mathop{\mathrm{argmin}}_{x\in \mathbb{R}^p} f(x) = \sum_{i=1}^n f_i(x),
\end{equation}
where $f_i(x) = \frac{1}{2} \|B_{(i)} x-b_{(i)}\|_2^2+\lambda_i \|x\|_q^q, B_{(i)} \in \mathbb{R}^{m_i\times p}, b_{(i)} \in \mathbb{R}^{m_i}$ for $i=1, \ldots,n,$ $\|x\|_q^q = \sum_{i=1}^p |x_i|^q$ for $0<q < 1$, and when $q=0,$ $\|x\|_q^q$ denotes the number of nonzero components of $x$.
Similar to Subsection 4.2, we take $n=10$, $p=256$, and $m_i = 150$ for $i=1,\ldots,n$. We take different $q = 0,1/2,2/3$ since their proximity operators have explicit forms and can be easily computed. In all cases, $\lambda_i = 0.5$ for each agent $i$ and four different step sizes are used. The experiment results are illustrated in Fig. {\ref{Fig:DLqLS}}.

By Fig. {\ref{Fig:DLqLS}}, the optimal step sizes for $q=0, 1/2$ and $2/3$ are $0.035, 0.012$ and 0.04, respectively. With these proper step sizes, when $q=0$, PG-ExtraPush performs linearly convergent, while for both $q=1/2$ and $2/3$, PG-ExtraPush decays sublinearly at the first several iterations, and then performs linearly. For these nonconvex cases, smaller step sizes generally imply the slower convergence.

\section{Conclusion}
\label{sc:conclusion}

In this paper, we propose a decentralized algorithm called PG-ExtraPush,
for solving decentralized composite consensus optimization problems over directed networks.
The algorithm uses a fixed step size and the proximal map of the nonsmooth part.
We show that with an appropriate step size, PG-ExtraPush converges to an optimal solution at a linear rate under some regular assumptions.
The effectiveness of PG-ExtraPush is also demonstrated by a series of numerical experiments. Specifically, PG-ExtraPush converges linearly and is significantly faster than Subgradient-Push, even the latter uses a hand-optimized step size. Moreover, we can observe from the numerical results that P-ExtraPush can generally accept a larger range of step size than PG-ExtraPush. Similar phenomenon between P-EXTRA and PG-EXTRA is also observed and verified in \cite{Shi-PGEXTRA2015}. Moreover, we show the potential of PG-ExtraPush for solving the decentralized nonconvex regularized optimization problems. In such nonconvex cases, PG-ExtraPush performs an eventual linear rate, i.e., the algorithm decays linearly starting from several iterations but not the initial iteration. However, its convergence as well as the rate of convergence in the general convex and nonconvex cases have not been studied in this paper, and we will investigate them in the future.

\end{document}